\documentclass[journal]{IEEEtran}

\usepackage[x11names]{xcolor}
\usepackage{graphicx} 
\usepackage{amsmath,amsfonts}
\usepackage{array}
\usepackage{textcomp}
\usepackage{stfloats}
\usepackage{url}
\usepackage{verbatim}
\usepackage{graphicx}
\usepackage{enumitem}
\usepackage{tikz}
\usepackage{multirow}
\usepackage{tabularx}
\usepackage{listings}
\usepackage{makecell}
\usepackage{algpseudocode}
\usepackage{siunitx} 
\usepackage{mathtools}
\usepackage{amsthm}
\usepackage{float}
\usepackage{comment}
\usepackage{hyperref} 
\usepackage[ruled,lined,boxed,linesnumbered]{algorithm2e}
\usepackage{balance}
\usepackage{booktabs} 
\usepackage{lipsum}
\usepackage[caption=false,font=footnotesize]{subfig}
\usepackage[table]{xcolor}
\onecolumn   

\clearpage

\newpage
\clearpage
\twocolumn

\title{Bridging the Initialization Gap: A Co-Optimization Framework for Mixed-Size Global Placement}
\author{
Yuhao Ren\IEEEauthorrefmark{1}\thanks{*These authors contributed equally to this work.},~\IEEEmembership{Student Member, IEEE},
Yiting Liu\IEEEauthorrefmark{1},~\IEEEmembership{Member, IEEE},
Yanfei Zhou,~\IEEEmembership{Student Member, IEEE},
Zhiyu Zheng,~\IEEEmembership{Student Member, IEEE},
Li Shang,~\IEEEmembership{Member, IEEE},
Fan Yang,~\IEEEmembership{Member, IEEE},
Zhiang Wang,~\IEEEmembership{Member, IEEE}

\thanks{The authors Yuhao Ren, Yanfei Zhou, Zhiyu Zheng, 
Fan Yang, and Zhiang Wang are with the State Key Lab of Integrated Circuits and Systems, College of Integrated Circuits and Micro-Nano Electronics, Fudan University, Shanghai. Yiting Liu and Li Shang is with the College of Computer Science and Artificial Intelligence, Fudan University, Shanghai.}}

\begin{document}

\maketitle

\begin{abstract}
Global placement is a critical step with high computation complexity in VLSI physical design. Modern analytical placers formulate the placement problem as a nonlinear optimization, where initialization highly affects both convergence and final placement quality.
However, existing initialization methods exhibit a trade-off: \emph{area-aware initializers} account for cell areas but are computationally expensive and often dominate total runtime, while fast \emph{point-based initializers} ignore cell area, leading to a modeling gap that impairs convergence and solution quality. 

This paper proposes a lightweight co-optimization framework that bridges this initialization gap through two strategies: 
(i) an \emph{area-hint refinement initializer}, which incorporates heuristic cell area information into a signed graph signal by augmenting the netlist graph with virtual nodes and negative-weight edges, yielding an area-aware and spectrally smooth placement initialization; and (ii) a \emph{macro-schedule placement} progressively restores area constraints, enabling a smooth transition from the refined initializer to the full area-aware objective and yielding high-quality placement results.
We evaluate the proposed framework using macro-heavy academic benchmarks ISPD2005 and two real-world industrial designs across two technology nodes (12 cases total). Experimental results show that our framework consistently improves half-perimeter wirelength (HPWL) over point-based initializers in 11/12 cases, achieving up to 2.2\% HPWL reduction, while running significantly faster ($\sim$100$\times$) than the state-of-the-art area-aware initializer~\cite{chen2023placement}. 

\end{abstract}

\section{Introduction}

\label{sec:intro}
Global placement determines the locations of all movable objects within the chip layout. It minimizes total wirelength while satisfying density, fixed I/O pins, placement blockages, and region boundary constraints.
Modern analytical placers formulate the placement problem as a nonlinear optimization problem that combines a convex wirelength objective and an electrostatics-based non-convex density function~\cite{eplace,replace,gu2020dreamplace}. 
This formulation scales efficiently to large designs and is typically solved using iterative gradient-based optimization. 
The placement performance is sensitive to the \emph{initial solution}: a good starting point can lead to faster convergence and higher-quality placement solutions~\cite{eplace}.

\begin{figure*}[!t]
  \centering
  \includegraphics[width=0.75\textwidth]{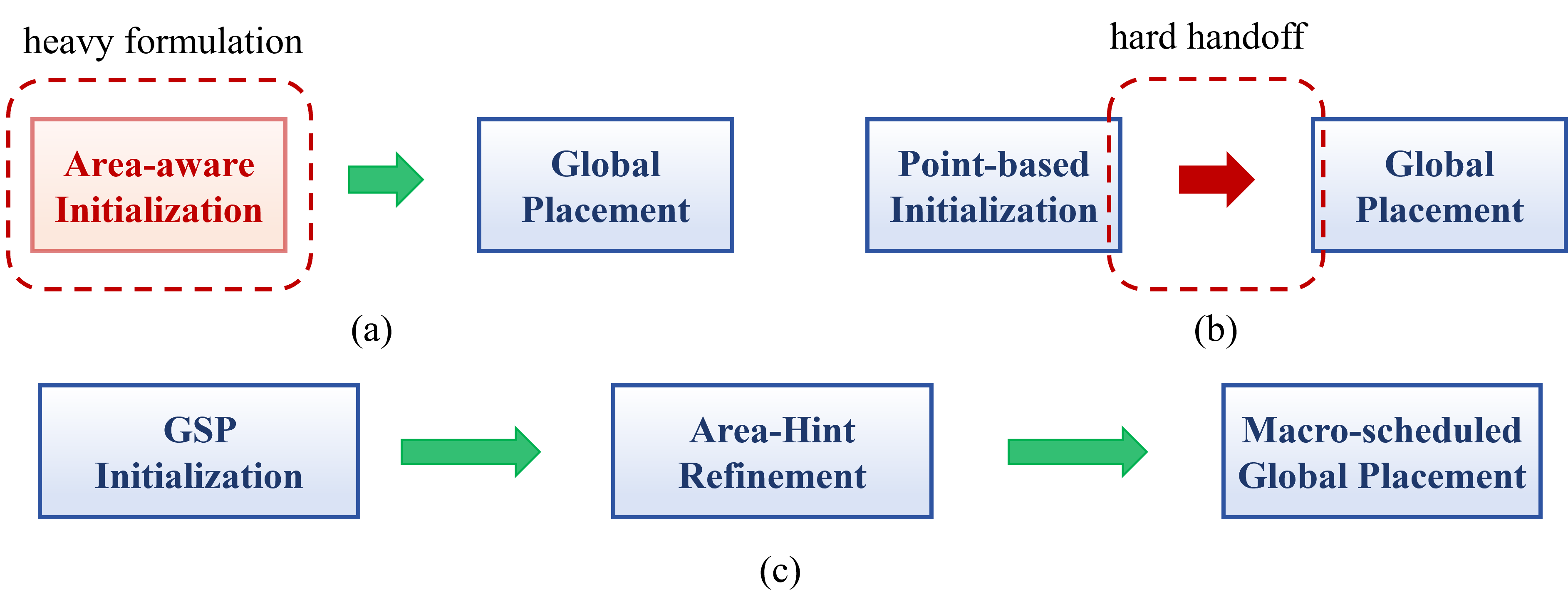} 
  \caption{
  (a) QCQP initialization, 
  (b) GSP initialization, and 
  (c) Our co-optimization framework.}
  \label{fig:framework}
  \vspace{-0.5em} 
\end{figure*}

Existing initialization methods can be broadly classified into two categories based on how they represent placement objects: \emph{area-aware initializers} and \emph{point-based initializers}. 
\textbf{Area-aware initializers} explicitly model the size and shape of each placement object, enforce density constraints, and penalize overlap.
By aligning with the analytical placer objectives, they produce physically realistic starting points but at a high computational cost (Figure~\ref{fig:framework}(a)).
Chen et al.~\cite{chen2023placement} formulate initialization as a sphere-constrained Quadratically Constrained Quadratic Program (QCQP), solved by sequential subspace optimization and reweighting.
Although this approach improves post-detailed-placement HPWL, its runtime grows rapidly with design size and may exceed that of the main global placement stage. 
Yao et al.~\cite{yao2025LLMevolution} explore large language models (LLMs) to evolve placement heuristics within DREAMPlace, adjusting initialization, preconditioning and gradient updates. 
However, this method only tunes numerical strategies rather than redesigning the underlying optimization model, and the heavy LLM reasoning time dominates total runtime.

In contrast, \textbf{point-based initializers} abstract each object as a zero-area point. This simplification enables lightweight computation and good scalability, making such methods widely used in modern placement flows. 
Early quadratic-wirelength approaches~\cite{chen2008ntuplace3, eplace} neglected density and overlap, producing rough but fast initial solutions.
DREAMPlace~\cite{gu2020dreamplace} further simplified initialization by randomly placing all cells within a small central region.
Recently, GiFt~\cite{liu2024power} provides a more advanced point-based initializer based on Graph Signal Processing (GSP). 
It efficiently extracts the global structure of the netlist by smoothing graph signals, achieving good performance on standard-cell–only designs.
However, GiFt models all cells as zero-area points and does not explicitly consider density constraints. This simplification degrades placement performance on macro-heavy mixed-size designs.
Therefore, while point-based initializers are fast and scalable, their lack of area awareness leads to a gap between initialization and the physically realistic analytical placer (Figure~\ref{fig:framework}(b)).


We argue that initialization should be a lightweight yet topology-revealing process that captures the global structure of the design and provide a coherent starting point for analytical optimization.
Instead of treating initialization and placement as two disjoint stages, we propose a co-optimization framework that smoothly transitions from a fast, topology-preserving point model to a full area-aware analytical model.
As illustrated in Figure~\ref{fig:framework}(c), the proposed framework bridges the initialization gap through two coordinated strategies:
\begin{itemize}[noitemsep,topsep=0pt,leftmargin=*]
\item \textbf{Area-hint refinement}: A lightweight initializer augments the netlist with virtual nodes and signed edges to encode approximate cell area information into the graph signal. This produces an area-aware yet spectrally smooth initialization that preserves GiFt’s efficiency while improving area awareness;
\item \textbf{Macro-scheduled placement}: The downstream global placement is enhanced into a macro-scheduled optimization that progressively restores macro footprints. Inspired by ITOP~\cite{tunneling}, this strategy enables a continuous and physically consistent transition from the refined initialization to the full area-aware placement objective. Unlike ITOP’s abrupt blockage relaxation, our macro tunneling gradually restores macro areas, ensuring stability and high-quality placement results.
\end{itemize}
Together, these strategies form a unified co-optimization flow that combines the speed of point-based initialization with the accuracy of area-aware placement, effectively bridging the initialization gap and producing high-quality placement results with high efficiency.
The key contributions of this paper are as follows:
\begin{itemize}[noitemsep,topsep=0pt,leftmargin=*]
    \item We present a co-optimization framework that bridges the gap between initialization and global placement by coupling a GSP-based point model with an area-aware analytical placer through a smooth transition.
    
    \item We encode area hints directly on the netlist graph by inserting virtual nodes and negative edges, and we introduce a signed-graph spectral filtering scheme that preserves global structure while modeling repulsive relations.

    \item We implement a macro-scheduled global placement that progressively restores macro footprints to ensure continuity from initialization to global placement optimization.
    
    \item On the ISPD2005 benchmarks and two real-world designs across two technology nodes (12 cases total), our method outperforms DREAMPlace  on 11/12 cases, achieving up to 2.2\% HPWL reduction, while running significantly faster ($\sim$100$\times$) than the state-of-the-art area-aware initializer.


    \item  Across different density targets, our initializer consistently achieves lower HPWL than DREAMPlace, demonstrating robustness to density parameters and random seeds.
    
\end{itemize}


The rest of this paper is organized as follows. 
Section II describes preliminaries for the rest of this paper. 
Section III presents the overview of our approach. Section IV-VI describe the three main components: GSP-based initialization, area-hint refinement and macro-scheduled global placement. Section VII introduces the autotuner. Section VIII presents the experimental results. 
We conclude the paper in Section IX.

\section{Preliminaries}

\begin{table}[hbtp]
\centering
\caption{Terminology and notation.}
\label{tab:notation}
\scriptsize
\renewcommand{\arraystretch}{1.05}
\setlength{\tabcolsep}{2pt}
\begin{tabular}{@{}|p{0.28\linewidth}|p{0.70\linewidth}|@{}}
\hline
\textbf{Notation} & \textbf{Description} \\
\hline
$G=(V,E)$ & Weighted undirected graph.\\
$A=[w_{i,j}]$ & Adjacency matrix; $w_{i,j}$ is the weight of edge $(v_i,v_j)\!\in\!E$.\\
$D$ & Degree matrix, $D_{ii}=\sum_j w_{i,j}$.\\
$L=D-A$ & Laplacian matrix of a graph.\\
$\widetilde{L}=D^{-1/2} L D^{-1/2}$ & Normalized Laplacian matrix.\\
$g\in\mathbb{R}^N$ & Graph signal; $g_i$ is the value at node $v_i$.\\
$\widehat{g}=U^\top g$ & Graph Fourier transform (GFT); $\widehat{g}_k=\langle \mathbf{u}_k,g\rangle$.\\
$S(g)$ & Smoothness of graph signal: $S(g)=\sum_{(i,j)\in E} w_{i,j}(g_j-g_i)^2$.\\
$\mathrm{zc}(g)$ & Edge zero-crossing count $\bigl|\{(v_i,v_j)\in E:\; g_i g_j<0\}\bigr|$.\\
$H=U\,h(\Lambda)\,U^\top$ & Graph signal filter with spectral response $h(\lambda)$.\\
\hline
$\mathbf{r}_k=(x_k,y_k)$ & Coordinates of movable object $k$.\\
$a_k$ & The area of movable object $k$.\\
$W(\mathbf{r})$ & Smooth wirelength model (WA).\\
$D(\mathbf{r})$ & Density penalty (overflow cost).\\
$\gamma$ & WA smoothing parameter.\\
$\bar{x}_e^{\pm},\;\bar{y}_e^{\pm}$ & WA soft extrema for net $e$ along $x$/$y$.\\
$\rho(\mathbf{r})$ & Continuous density.\\
$\Pi(\cdot)$ & A smooth box kernel that spreads each object onto nearby bins.\\
$\rho_{\mathrm{tgt}}$ & Target density.\\
$\phi$ & Electrostatic potential.\\
 $\mathbf{E}=\nabla\phi$ & Electric field.\\
$\Omega$ & Placement region; $B_b$ is bin region $b$; $\rho_b$ is bin density.\\
$\varepsilon$ & Poisson-equation scaling constant.\\
\hline
$L_{\text{hint}}$ & Laplacian matrix after embedding area hints.\\
$w_m,h_m$ & Macro width and height.\\
$x_m,y_m$ & Coordinates of macro center.\\
$\Delta x_{i,m},\Delta y_{i,m}$ & Offsets from instance $i$ to macro $m$ center.\\
$\eta$ & Macro charge–density control (scheduled over iterations); controls spread/concentration of the macro’s Gaussian density model. \\
$\rho_\eta(\Delta x,\Delta y)$ & Truncated, area-normalized 2D Gaussian (macro charge).\\
$\rho_t(\Delta x,\Delta y)$ & Time-scheduled charge density at iteration $t$.\\
$\alpha = t/T$ & Normalized iteration index; $T$ is the total number of iterations. \\
\hline
\end{tabular}
\end{table}


In this section, we first review the mathematical foundations of graph signal processing (Section~II-A), which underpin the derivations used in our GSP-based initialization (Section~IV). 
We then present the electrostatics-based placement formulation (Section~II-B), which serves as the backbone of our macro-scheduled global placement(Section~VI). 
Table~\ref{tab:notation} summarizes the key symbols and their meanings.

\subsection{Graph Signal Processing}
Given a weighted undirected graph $G=(V,E)$ with node set $V$ and edge set $E$,  
it can be represented by an adjacency matrix $A = [w_{i,j}]\in\mathbb{R}^{N\times N}$, where
\[
w_{i,j} =
\begin{cases}
weight~of~edge_{i,j}, & \text{if }(v_i,v_j)\in E\\
0,  & \text{otherwise.}
\end{cases}
\]
The graph Laplacian matrix is defined as $L = D - A$,
where $D = \operatorname{diag}(d_1, d_2, \dots, d_N)\in\mathbb{R}^{N\times N}$ represent the degree matrix of $A$.
The normalized graph Laplacian matrix is $\widetilde{L} = D^{-\frac12}\,L\,D^{-\frac12}$.

A graph signal $g$ is a mapping $g\colon V \to \mathbb{R}$,
which can be written as an $N$-dimensional vector $g = \bigl[g_1,\,g_2,\,\dots,\,g_N\bigr]^\top$, and each $g_i$ encodes information associated with node $v_i$~\cite{shuman2013emerging}.  
Graph Signal Processing (GSP) is the field dedicated to the analysis and manipulation of such signals on graphs, intending to extract meaningful insights.

\subsubsection{Smoothness}
In the context of graph signal processing (GSP), a graph signal is considered smooth if the signal values associated with connected nodes tend to be similar~\cite{dong2016learning}.
The smoothness of the graph signal can be quantified via the graph Laplacian quadratic form:
\begin{equation}\label{eq:smoothness}
S(g)
=\frac{1}{2}\sum_{v_i\in V}\bigl\|\nabla_i g\bigr\|_2^2
=\sum_{(v_i,v_j)\in E}w_{i,j}\,(g_j - g_i)^2
\end{equation}

\subsubsection{Graph Fourier Transform}
The Graph Fourier Transform (GFT) of a graph signal $g$ is given by the projection of $g$ onto the eigenbasis of the graph Laplacian matrix. 
Since $L$ is real and symmetric, it admits an eigendecomposition 
$L = U \,\Lambda\, U^\top$,
where
$\Lambda = \operatorname{diag}(\lambda_0,\lambda_1,\dots,\lambda_{n-1})$
is the diagonal matrix of eigenvalues (``frequencies'')
satisfying
$\lambda_0 \le \lambda_1 \le \cdots \le \lambda_{n-1}$,
and
$U = [\mathbf{u}_0,\mathbf{u}_1,\dots,\mathbf{u}_{n-1}]$
is an orthonormal matrix whose columns are the corresponding eigenvectors.

The \emph{forward GFT} of a signal $g\in\mathbb{R}^n$ is defined as
\begin{equation}\label{eq:GFT}
    \widehat{g} \;=\; U^\top g,
\end{equation}
so that the $k$-th spectral coefficient
$\widehat{g}_k = \langle \mathbf{u}_k,\,g \rangle$
quantifies the component of $g$ oscillating at “frequency” $\lambda_k$.

The \emph{inverse GFT} reconstructs the original signal via
\begin{equation}\label{eq:invert-GFT}
    g \;=\; U\,\widehat{g}
    \;=\;\sum_{k=0}^{n-1}\widehat{g}_k\,\mathbf{u}_k.
\end{equation}

This spectral framework generalizes the classical Fourier transform to irregular domains, enabling analysis and filtering of graph-supported data in the spectral domain.

\subsubsection{Spectral Domain}
From the graph-spectral viewpoint, the graph Fourier transform (GFT) expands a graph signal $g$ in the Laplacian eigenbasis $\{u_k\}$ (Eq.~\ref{eq:invert-GFT}). We define the edge zero-crossing count as
\begin{equation}
  \mathrm{zc}(g)\;=\;\bigl|\{(i,j)\in E:\; g_i\,g_j<0\}\bigr| ,
\end{equation}
which means the number of edges whose endpoints have opposite signs. Eigenvectors associated with larger eigenvalues \(\lambda_k\) oscillate more across edges and thus exhibit more zero crossings~\cite{shuman2013emerging}, which correspond to higher graph frequencies and are less smooth. 
Therefore, a smooth signal concentrates energy on low-$\lambda$ components: $|\hat{g_k}|$ should be relatively large for small $\lambda_k$ and surppressed for large $\lambda_k$.

\subsubsection{Graph Filter}
A graph filter is a linear operator that modifies a graph signal by selectively amplifying or attenuating its spectral components, which is often defined as:
\begin{equation}\label{eq:filter}
    H = U\,h(\Lambda)\,U^\top,
\end{equation}
where
$h(\Lambda) = \operatorname{diag}\bigl(h(\lambda_0),\,h(\lambda_1),\,\dots,\,h(\lambda_{n-1})\bigr)$
is a spectral response function prescribing the gain $h(\lambda_k)$ at graph frequency~$\lambda_k$.
Applied to a graph signal \(g\in\mathbb{R}^n\), the filtered signal is
$g_{\mathrm{out}} = H\,g = U\,h(\Lambda)\,U^\top\,g$,
which in the spectral domain corresponds to elementwise multiplication
$\widehat{g}_{\mathrm{out},k} = h(\lambda_k)\,\widehat{g}_k.$
This framework generalizes classical filtering to arbitrary graph domains, allowing tasks such as smoothing, denoising, and band-pass filtering of graph-structured data.

\subsection{Electrostatics-Based Placement}

Electrostatics-Based Placement formulates global placement as a smooth, unconstrained optimization that minimizes wirelength while enforcing near-uniform bin density. 
Let movable objects (cells/macros) be indexed by $k$ with coordinates $\mathbf{r}_k=(x_k,y_k)$. 
The objective is
\begin{equation}
    \min_{\{\mathbf{r}_k\}} F(\mathbf{r}) \;=\; W(\mathbf{r}) \;+\; \lambda\, D(\mathbf{r}),
\end{equation}
where $W$ is a smooth wirelength model (e.g., WA/LSE)~\cite{hsu2011tsv, naylor2001non} and $D$ penalizes density overflow; $\lambda>0$ balances the two terms.

\subsubsection{Wirelength model} ePlace uses the \emph{weighted-average} (WA) wirelength~\cite{hsu2011tsv}, applied per net and per axis, which smoothly approximates HPWL without introducing large smoothing bias. 
For a net $e$ with pin coordinates $\{x_i\}_{i\in e}$, define
\begin{equation}
    \bar{x}_e^{+}=\frac{\sum_{i\in e} x_i\, e^{x_i/\gamma}}{\sum_{i\in e} e^{x_i/\gamma}},\qquad
    \bar{x}_e^{-}=\frac{\sum_{i\in e} x_i\, e^{-x_i/\gamma}}{\sum_{i\in e} e^{-x_i/\gamma}},
\end{equation}

and analogously $\bar{y}_e^{\pm}$. The WA wirelength is
\begin{equation}
    W(\mathbf{r}) \;=\; \sum_{e\in\mathcal{N}}
    \Big[(\bar{x}_e^{+}-\bar{x}_e^{-})+(\bar{y}_e^{+}-\bar{y}_e^{-})\Big],
\end{equation}
with smoothing parameter $\gamma$. 
The gradients $\partial W/\partial x_i$ and $\partial W/\partial y_i$ drive instances on the same net to move closer to one another, thereby reducing wirelength.

\subsubsection{Electrostatics-based density} The placement core is discretized into bins. 
Let the continuous density be
\begin{equation}
    \rho(\mathbf{r}) \;=\; \sum_k a_k\, \Pi(\mathbf{r}-\mathbf{r}_k),
\end{equation}
where $a_k$ is object area and $\Pi$ is a smooth box kernel that spreads each object onto nearby bins. 
ePlace interprets density spreading through electrostatics: compute potential $\phi$ by solving the Poisson equation
\begin{equation}
    \nabla^2 \phi(\mathbf{r}) \;=\; -\,\frac{\rho(\mathbf{r})-\rho_{\text{tgt}}}{\varepsilon},
    \qquad
    \left.\frac{\partial \phi}{\partial n}\right|_{\partial \Omega}=0,
\end{equation}
where $\rho_{\text{tgt}}$ is the target density and $\varepsilon$ is a scaling constant. 
The electric field $\mathbf{E}=\nabla\phi$ induces \emph{density forces} on objects that push mass from overfull to underfull regions. 
The density penalty can be written as an $L_2$ mismatch,
\begin{equation}
    D(\mathbf{r}) \;=\; \frac{1}{2}\int_{\Omega}\big(\rho(\mathbf{r})-\rho_{\text{tgt}}\big)^2\,d\mathbf{r},
\end{equation}
whose gradient w.r.t.\ $\mathbf{r}_k$ is proportional to $a_k\,\mathbf{E}(\mathbf{r}_k)$. 
The Poisson solve and its gradients are evaluated efficiently by an FFT-based spectral solver, yielding near $O(M\log M)$ cost for $M$ bins.

\section{Overview of Our Approach}

As shown in Figure~\ref{fig:framework}(c), our approach consists of three major steps.

\noindent
\textbf{GSP-Based Initialization} (Section~IV): During this step, each instance (macro or standard cell) is treated as a zero-area point. An instance graph is induced from the netlist, and the instance coordinates are modeled as node-level graph signal.
We then apply a low-pass (small-eigenvalue) spectral filter to smooth this graph signal, drawing strongly connected instances closer in coordinate space while simultaneously extracting salient global structure in the spectral domain by emphasizing low-frequency components.

\noindent
\textbf{Area-Hint Refinement} (Section~V): During this step, we inject footprint information
by augmenting the instance graph with virtual nodes and signed edges, and design a signed-graph spectral filter for the resulting graph signal. Virtual nodes encode area hints (e.g., 
macro footprints, bin capacities), while negative edges model repulsive relations that 
discourage overcrowding near macros and within overfull bins. Filtering on this signed graph 
emulates area effects during initialization, reducing overlaps between cells and macros and alleviating standard-cell overconcentration. This step yields a smoother handoff from the GSP point-based model to the area-aware analytical placer.

\noindent
\textbf{Macro\mbox{-}Scheduled Global Placement} (Section~VI): We retain the electrostatics-based placement framework and introduce two scheduling strategies. In both, each macro is treated as a nonuniform, iteration-dependent charge distribution. 
(i) \emph{Charge-redistribution schedule}: each macro is modeled as a 2D Gaussian density source with a scheduled variance $\sigma^2(t)$ that monotonically increases over iterations; the total charge (mass) is conserved, so the macro’s charge (mass) gradually disperses into neighboring bins. 
(ii) \emph{Charge-restoration schedule}: we progressively restore the macro’s charge from center to edges as iterations advance, emulating the gradual recovery of its hard footprint. 
These schedules ensure a smooth transition from initialization to global placement and mitigate the abrupt density cliffs of hard-macro models, which would otherwise create high potential barriers that trap standard cells in local valleys.

Section~\ref{sec:GSP-Based Initialization} provides the details of the GSP-based initialization; Section~\ref{sec:Area-Hint} presents the area-hint refinement; and Section~\ref{sec:macro-schedule} elaborates on the macro\mbox{-}scheduled global placement.

\section{GSP-Based Initialization}
\label{sec:GSP-Based Initialization}

A placement instance can be formulated as a graph $G=(V,E)$, where $V$ is the set of objects (e.g., I/Os, macros, and standard cells) and $E$ is the set of nets. The main objective is to find a placement 
$g = \bigl[(x_1,y_1),\,(x_2,y_2),\,\dots,\,(x_N,y_N)\bigr]^\top \in \mathbb{R}^{N\times2}$
that minimizes the total wirelength $S(g)$ subject to density constraints (i.e., the density $\rho_b(g)$ of each bin $b$ does not exceed a target $\rho_{tgt}$):
\begin{equation}
\min_g f(g) = S(g)
\quad\text{s.t.}\quad
\rho_b(g) \;\le\; \rho_{tgt}.
\end{equation}

The total wirelength $S(g)$ is estimated as the weighted sum of squared distances between connected objects:
\begin{equation}\label{eq:Wirelength objective}
S(g)
=\sum_{(v_i,v_j)\in E} w_{i,j}\bigl[(x_i - x_j)^2 + (y_i - y_j)^2\bigr].
\end{equation}

\begin{figure}[!b]
  \centering
  \includegraphics[width=\columnwidth]{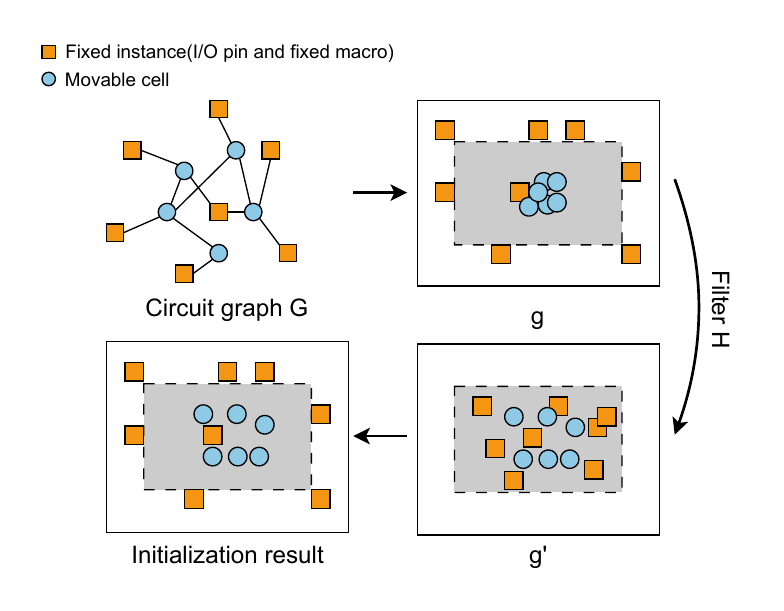}
  \caption{GSP-based initialization flow}
  \label{fig: GSP-based initialization}
\end{figure}

As demonstrated in Eq.~\ref{eq:smoothness} and Eq.~\ref{eq:Wirelength objective}, the optimization objective in placement, which aims to minimize quadratic wirelength, aligns with the goal of enhancing graph signal smoothness from the perspective of GSP. 
Following Section II-A(c), we smooth the graph signal by suppressing the spectral coefficients $\hat{g}_k$ associated with large eigenvalues $\lambda_k$. 
Accordingly, the core of a GSP-based initialization is to construct a filter matrix $H$ (Eq.~\ref{eq:filter}) whose response preserves the spectral coefficients associated with small eigenvalues (good passband) while strongly attenuating those associated with large eigenvalues~\cite{liu2024power}.

Liu. etc~\cite{liu2025gsp-based} has proved the normalized adjacent matrix $\widetilde{A} = D^{-\frac{1}{2}}AD^{-\frac{1}{2}}$ is a graph filter corresponding to the  filter function, that is,
\begin{equation}\label{eq:GiFt}
    \tilde{A}=U \Lambda_{A} U^{T}
=U\operatorname{diag}\!\big(1-\lambda_{1},\,1-\lambda_{2},\,\ldots,\,1-\lambda_{n}\big)\,U^{T}.
\end{equation}
Because normalized Laplacian \(\widetilde{L}\) has eigenvalues confined to \([0,2]\), the baseline operator \(\tilde{A}\) with response \(h(\lambda)=1-\lambda\) tends to suppress mainly mid–frequency components, which is not ideal for suppressing $\hat{g_k}$ for large $\lambda_k$

To strengthen suppressing, two amendments can be made. 
First, augment the adjacency with self-loops,
\begin{equation}
    A_\sigma \;=\; A + \sigma I,\qquad \sigma \ge 0,
\end{equation}

which rescales the spectrum so that large eigenvalues shrink~\cite{wu2019simplifying}; the corresponding augmented operator behaves more like a low-pass filter.

Second, generalize the augmented operator by a spectral power, using \(\tilde{A}_\sigma^{\,k}\). This changes the eigenvalue response from
\(h(\lambda)=1-\lambda\) to
\begin{equation}
    h_k(\lambda) \;=\; (1-\lambda)^k,
\end{equation}
where \(k\) controls the filter strength: larger \(k\) yields more pronounced attenuation of high-frequency content. 
By tuning \(\sigma\) and \(k\), one can regulate the degree of high-frequency suppression and obtain smooth signals at multiple resolutions.

To avoid over globally smooth signals overlook valuable local information, which can be characterized as globally high-frequency yet locally smooth~\cite{liu2023personalized}, a multi-eigenvalue graph filters $H$ is built as follow:
\begin{equation}\label{eq:GiFt filter}
    H = \alpha_{low}\widetilde{A}^{low\_k}_{low\_\sigma} + \alpha_{mid}\widetilde{A}^{mid\_k}_{mid\_\sigma} + \alpha_{high}\widetilde{A}^{high\_k}_{high\_\sigma}
\end{equation}

 The three bands are nstantiated by
\(\tilde{A}^{2}_{2}\) (high-pass), \(\tilde{A}^{2}_{4}\) (mid-pass), and \(\tilde{A}^{4}_{4}\) (low-pass)~\cite{liu2024power}. 
The high-pass \(\tilde{A}^{2}_{2}\) intentionally retains a portion of high-frequency components to capture local details; 
the mid-pass \(\tilde{A}^{2}_{4}\) emphasizes intermediate frequencies; 
and the low-pass \(\tilde{A}^{4}_{4}\) predominantly preserves low-frequency content, thereby promoting global smoothness. 
The weights \(\alpha_{\text{high}}, \alpha_{\text{mid}}, \alpha_{\text{low}}\) govern the trade-off between locally smooth (detail-preserving) and globally smooth behavior. All these parameters can be fine-tuned by AutoDMP latter(section III-D)

Once the filter $H$ is constructed, we perform GSP-based initialization as Figure~\ref{fig: GSP-based initialization}: 1) Initialize the coordinate signals by sampling coordinates for movable instances within the placement region, while pinning all fixed objects (e.g., fixed I/Os and fixed macros) at their predetermined locations, thereby yielding the initial graph signal $g$. 
2) Apply the graph filter $H$ to graph signal $g$ to obtain the smoothed signal graph $g'$. 
3) Re-enforce fixed-location constraints by overwriting the coordinates of fixed objects with their predetermined locations, ensuring the filtering step has not perturbed them.

\begin{figure}[b]
  \centering
  \includegraphics[width=\columnwidth]{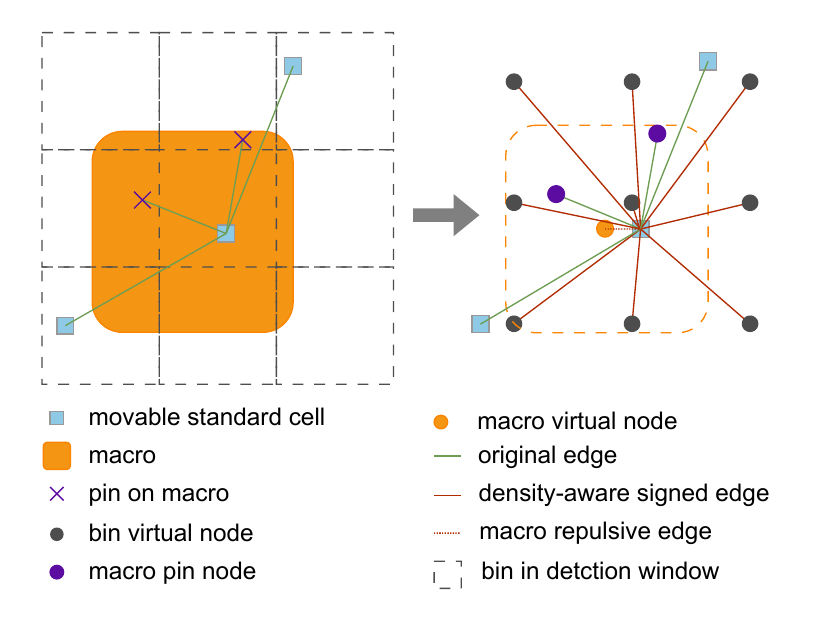}
  \caption{Area-hint integration}
  \label{fig: areahint}
\end{figure}

\section{Area-Hint Refinement}
\label{sec:Area-Hint}

Studies on signed graphs~\cite{cheung2018robust,furutani2019graph} show that negative edge weights encode dissimilarity: when $\omega_{i,j}<0$, the signal values $g_i$ and $g_j$ are encouraged to diverge(i.e, $|g_i-g_j|$ should be large). 
Incorporating such pairwise repulsion into the graph benefits GSP-based modeling and filtering. 
Accordingly, we augment the instance graph with virtual nodes and negative edges to inject area hints: virtual nodes represent repulsive sources (fixed macros or overfull bins), and negative edges connect them to nearby standard cells, discouraging proximity to fixed macros and high-density regions. 
In this way, area hints are embedded into the graph signal and exploited by our signed-graph filter.

To embed area hints into the graph signal we employ three constructions as shown in Figure~\ref{fig: areahint}.

\noindent \textbf{Macro pin expansion.}
Instead of modeling a macro as a single fixed node, we split it into multiple fixed nodes placed at its pin sites, which better reflects the macro's spatial influence.

\noindent \textbf{Macro virtual node with repulsive (negative) edges.}
For each macro \(m\), introduce a virtual fixed node \(v_m\).
Connect \(v_m\) to every movable instance \(i\) who overlaps macro \(m\)'s footprint by a \emph{negative-weight} edge. 
We use an axis-normalized distance ratio
\begin{equation}
  \textit{maxRatio}_{i,m}
  \;=\;
  \max\!\Bigl(
    \tfrac{\lvert \Delta x_{i,m} \rvert}{w_m/2},
    \tfrac{\lvert \Delta y_{i,m} \rvert}{h_m/2}
  \Bigr),
\end{equation}
where \(w_m,h_m\) are the macro width/height and \(\Delta x_{i,m},\Delta y_{i,m}\) are offsets from instance \(i\) to the macro center (along \(x\) and \(y\)). 
The edge weight is distance–decayed:
\begin{equation}
  w^{\mathrm{macro}}_{i,m}
  \;=\;
  -\, e^{-\textit{maxRatio}_{i,m}}\cdot \textit{baseWeight}_i,
\end{equation}
where \(\textit{baseWeight}_i>0\) is the mean weight of all original netlist edges incident to instance \(i\).

\noindent \textbf{Bin virtual node with density-aware signed edges.}
Partition the placement area into \(\mathrm{numBin}_x\times\mathrm{numBin}_y\) bins. 
For each bin \(b\), introduce a virtual node \(v_b\) and define
\begin{equation}
  \phi_b \;=\; 2\,\operatorname{sigmoid}\!\big(\mu(\mathrm{D}_b-\mathcal{C})\big)\;-\;1 ,
\end{equation}
where $\mathrm{D}_b$ is the current density of bin \(b\), $\mathcal{C}$ is the bin capacity, and \(\mu>0\) controls the slope of the logistic mapping.
Then we connect every movable instance \(i\) within the detection window \(\mathcal{N}(b)\) to \(v_b\) by a signed edge with weight
\begin{equation}
  w^{\mathrm{bin}}_{i,b}
  \;=\;
  -\, e^{-\textit{maxRatio}_{i,b}}\;\phi_b\;\textit{baseWeight}_i ,
\end{equation}
where the axis-normalized distance ratio to bin \(b\) is
\begin{equation}
  \textit{maxRatio}_{i,b}
  \;=\;
  \max\!\Bigl(
    \tfrac{\lvert x_i - x_b \rvert}{w_b/2},
    \tfrac{\lvert y_i - y_b \rvert}{h_b/2}
  \Bigr),
\end{equation}
with \((x_b,y_b)\) the bin center and \(w_b,h_b\) the bin width/height. 
The detection window \(\mathcal{N}(b)\) is an \(n_x \times n_y\) neighborhood centered at \(b\), with defaults 
\(n_x=\max\{1,\lfloor 0.1\,\mathrm{numBin}_x \rfloor\}\) and 
\(n_y=\max\{1,\lfloor 0.1\,\mathrm{numBin}_y \rfloor\}\). 
This choice yields \(\phi_b>0\) for overfull bins and \(\phi_b<0\) for underfull bins, making \(w^{\mathrm{bin}}_{i,b}\) \emph{negative} (repulsive) in overfull regions and \emph{positive} (attractive) in underfull regions, thereby discouraging overcrowding and encouraging utilization of slack bins.

\

After embedding area hints, the resulting graph edges become \emph{signed}. Consequently, the standard GSP-based construction that normalizes the (nonnegative) adjacency matrix to build a filter is no longer applicable. We therefore design a new refinement filter
\begin{equation}\label{eq:refine-filter}
  H_{\mathrm{r}}
  \;=\;
  \Bigl(\frac{\lambda_{\mathrm{up}} I - L_{\mathrm{hint}}}{\lambda_{\mathrm{up}}}\Bigr)^{k},
\end{equation}
whose eigenvalue response is
\begin{equation}
  h_{\mathrm{r}}^{(k)}(\lambda)
  \;=\;
  \Bigl(1-\frac{\lambda}{\lambda_{\mathrm{up}}}\Bigr)^{k}.
\end{equation}
Here $\lambda_{\mathrm{up}}$ is an upper bound on the spectrum of $L_{\mathrm{hint}}$, estimated via the Gershgorin circle theorem~\cite{weisstein2003gershgorin}.




\begin{figure*}[!b]
  \centering
  \includegraphics[width=\textwidth]{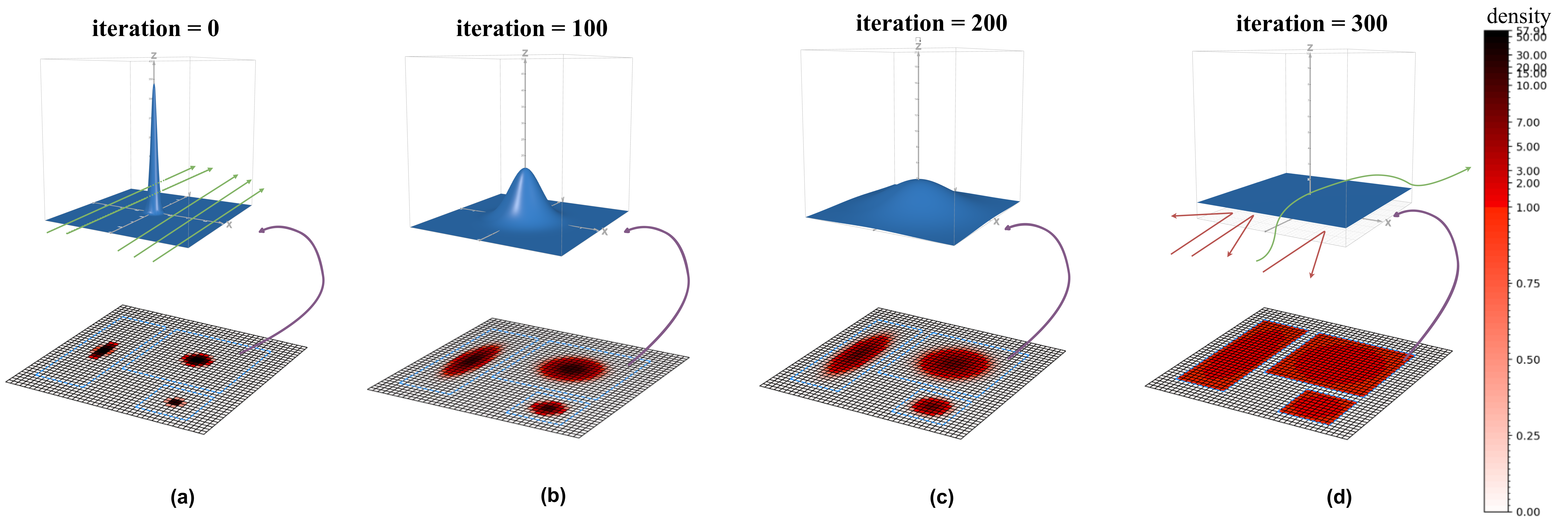}
  \caption{Macro schedule strategy: Panels (a)–(d) illustrate how the macro’s charge density evolves over iterations. In early iterations (a), the mass is highly concentrated near the center, creating a narrow barrier so standard cells can readily traverse the macro region; in later iterations (d), the distribution spreads to the full footprint and the barrier broadens, making traversal difficult.}
  \label{fig: Gaussion density}
  \vspace{-0.5em}
\end{figure*}

Applying the refinement filter $H_{\mathrm{r}}$ iteratively to the graph signals produced by the GSP-based initialization completes the \emph{area-hint refinement}. The updated initialization coordinates better respect macro footprints and bin capacities while preserving global (low-frequency) smoothness. As shown in Algorithm ~\ref{alg:area_hint}, we implement this step as a short fixed-point iteration: at iteration $k$, we (i) construct an area-hint Laplacian $L_{\mathrm{hint}}^{(k)}$ from the current placement (bin utilizations and macro masks), (ii) compute the corresponding refinement filter $H_{\mathrm{r}}^{(k)}$ from $L_{\mathrm{hint}}^{(k)}$, and (iii) update the graph signals. We repeat until a small iteration budget is reached.

\begin{algorithm}[ht]
\small
\caption{Iterative Area\mbox{-}Hint Refinement}
\label{alg:area_hint}
\SetKwInOut{Input}{Input}
\SetKwInOut{Output}{Output}
\Input{Initial graph signal $\mathbf{g}^{(0)}$ from GSP-based initialization; macro mask $\mathcal{M}$; bin capacities $\mathcal{C}$; iteration budget $T_{\mathrm{hint}}$; relaxation $\gamma \in [0,1]$.}
\Output{Refined graph signal $\mathbf{g}^{\star}$.}
\For{$k \leftarrow 0$ \KwTo $T_{\mathrm{hint}}-1$}{
    $L_{\mathrm{hint}}^{(k)} \gets BuildLap\!\left(\mathcal{M}, \mathcal{C}, \mathbf{g}^{(k)}\right)$\;
    $H_{\mathrm{r}}^{(k)} \gets ComputeFilter\!\left(L_{\mathrm{hint}}^{(k)}\right)$\;
    $\tilde{\mathbf{g}} \gets H_{\mathrm{r}}^{(k)}\,\mathbf{g}^{(k)}$\;
    $\mathbf{g}^{(k+1)} \gets (1-\gamma)\,\mathbf{g}^{(k)} + \gamma\,\tilde{\mathbf{g}}$\;
}
$\mathbf{g}^{\star} \gets \mathbf{g}^{(k+1)}$\;
\Return $\mathbf{g}^{\star}$\;
\end{algorithm}

\section{Macro-Schedule Global Placement}
\label{sec:macro-schedule}

After the area-hint refinement, we proceed to global placement. Unlike conventional flows, we gradually restore each macro’s footprint during optimization, thereby avoiding both the hard handoff between a point-model initialization and an area-aware placer and the early blocking of standard-cell motion by the high potentials around macros.

A naive schedule would directly scale macro areas by initializing each macro with a standard-cell–sized footprint and enlarging it across iterations, but this causes per-bin densities to change discontinuously between iterations, leading to abrupt jumps in the electrostatics-based potential and destabilizing the solver. 
To resolve this, we introduce a charge-distribution–based macro model that replaces abrupt area scaling.
Under the electrostatics analogy, a fixed macro can be viewed as a region with high, uniformly distributed charge density. 
Our scheduling scheme instead treats the macro’s “charge” as a non-uniform, iteration-dependent distribution, so the induced potential varies smoothly over iterations. 
This keeps the optimization well-conditioned while the macro footprint is progressively restored.

Next, we introduce two scheduling strategies; experiments for both are reported in Section~IV\mbox{-}C.
\subsection{Charge-redistribution}
We model each macro’s charge as an area–normalized, rectangularly truncated 2D Gaussian in \emph{local} coordinates
\(\Delta x_{i,m},\Delta y_{i,m}\) (offsets from the macro center) so that the integral over the original macro region equals its area \(wh\). That is:
\begin{equation}\label{eq: Gaussion_eta}
  \rho_{\eta}(\Delta x_{i,m},\Delta y_{i,m})=
  \frac{2\eta^{2}}{\pi\, \operatorname{erf}^{2}\!\!\left(\frac{\eta}{\sqrt{2}}\right)}
  e^{-2\eta^{2}\!\left(\frac{\Delta x_{i,m}^{2}}{w_m^{2}}+\frac{\Delta y_{i,m}^{2}}{h_m^{2}}\right)} .
\end{equation}
The support is restricted to \(|\Delta x_{i,m}|\le w_m/2\) and \(|\Delta y_{i,m}|\le h_m/2\) (and the density is zero outside this rectangle),
where \(\eta=\tfrac{w_m}{2\sigma_x}=\tfrac{h_m}{2\sigma_y}\) with \(\sigma_x,\sigma_y\) the Gaussian standard deviations.  
As \(\eta\to 0\), \(\rho_{\eta}\) becomes nearly flat over the macro and as \(\eta\to+\infty\), the mass concentrates at the center, which matches the “point \(\rightarrow\) area” progression.

To schedule the restoration over \(T\) iterations smoothly, let \(\alpha=t/T\in[0,1]\) and define two baseline controls :

\noindent\textbf{Scheme A: boundary–attenuation control.}
We control the \emph{edge decay ratio} \(r(\alpha)\in(0,1)\) (center-to-boundary amplitude) in a log-linear fashion,
\begin{equation}
  \log r(\alpha)=(1-\alpha)\log r_0+\alpha\log r_1,\;\
  \eta_A(\alpha)=\sqrt{-2\log r(\alpha)} .
\end{equation}
This produces a very \emph{flat tail} at the end (good spreading) but the initial concentration is weaker.

\noindent\textbf{Scheme B: FWHM control.}
We linearly schedule the normalized FWHM \(\beta(\alpha)\), defined as the FWHM divided by the macro width/height,
\begin{equation}
  \beta(\alpha)=\beta_{\min}+(\beta_{\max}-\beta_{\min})\,\alpha,\qquad
  \eta_B(\alpha)=\frac{\sqrt{2\ln 2}}{\beta(\alpha)} .
\end{equation}
This yields \emph{strong initial concentration} but the final spreading can be less uniform.

\noindent\textbf{Blended schedule.}
To combine the good early concentration of Scheme~B with the flat final spreading of Scheme~A, we interpolate smoothly using a “smoothstep’’ weight \(w(\alpha)\):
\begin{equation}
  w(\alpha)=
  \begin{cases}
    0, & \alpha\le \alpha_0,\\
    3z^2-2z^3,\ \ z=\dfrac{\alpha-\alpha_0}{\alpha_1-\alpha_0}, & \alpha_0<\alpha<\alpha_1,\\
    1, & \alpha\ge \alpha_1,
  \end{cases}
\end{equation}
and define the scheduled concentration parameter via geometric mixing
\begin{equation}\label{eq: eta_t}
  \eta(t)=
  \eta_B(\alpha)^{\,1-w(\alpha)}\,\eta_A(\alpha)^{\,w(\alpha)} .
\end{equation}
Combining Eqs.~\eqref{eq: Gaussion_eta} and \eqref{eq: eta_t}, the time-scheduled charge density is
\begin{equation}
  \rho_{t}(\Delta x_{i,m},\Delta y_{i,m})
  \;=\;
  \rho_{\eta(t)}(\Delta x_{i,m},\Delta y_{i,m}).
  \label{eq:rho_t}
\end{equation}

\subsection{Charge-restoration}
In the charge-redistribution strategy, early iterations may place high density near a macro’s center, and the resulting repulsive forces can push nearby cells excessively far, causing undesirable transient perturbations. To alleviate this, \emph{charge restoration} relaxes the requirement of strict mass conservation within the macro footprint and instead \emph{progressively restores} charge from the center toward the edges across iterations.

\begin{figure}[!t]
  \centering
  \includegraphics[width=\columnwidth]{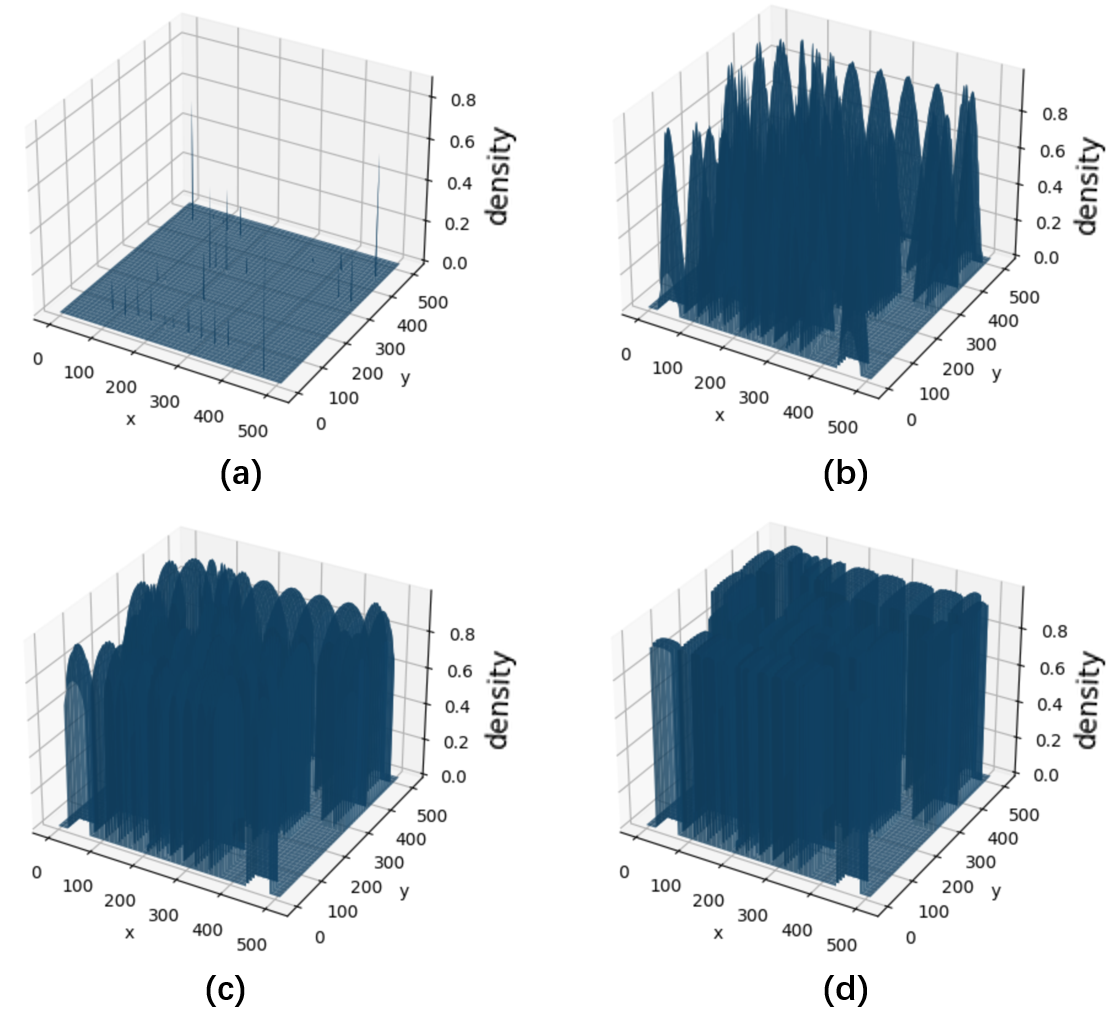}
  \caption{Macro\mbox{-}region density evolution under the charge\mbox{-}restoration strategy ($T=300$): (a) $t=0$; (b) $t=100$; (c) $t=200$; and (d) $t=300$.}
  \label{fig: exp density}
\end{figure}

We define a \textbf{Exponential-like model} whose attenuation toward the boundary is governed by an iteration-dependent scale parameter $\sigma>0$:
\begin{equation}
\label{eq: exp_sigma} 
\rho_{\sigma}(\Delta x_{i,m},\Delta y_{i,m})= e^{-\frac{{tan(\pi\frac{\Delta x_{i,m}}{w_m})}^2+{tan(\pi\frac{\Delta y_{i,m}}{h_m})}^2}{2\sigma^2}} . 
\end{equation}
which is smoothly bell-shaped near the center and rapidly decays to zero as $|\frac{\Delta x_{i,m}}{w_m}|$ or \(|\frac{\Delta y_{i,m}}{h_m}|\!\to\! 1\) due to the divergence of \(\tan(\cdot)\).
As \(\sigma\) increases, the restored high-density region expands outward from the center, emulating a center-to-edge ``fill-in'' of the macro footprint.
To make the expansion of the high\mbox{-}density region smoother, we schedule the scale parameter $\sigma$ as a function of iteration $t\!\in\!\{0,\ldots,T\}$:
\begin{equation}\label{eq:sigma_t}
  \sigma(t)
  \;=\;
  -\,\sigma_{\mathrm{factor}}\,T\,\ln\!\left(1-\frac{t}{T}\right),
\end{equation}
where $\sigma_{\mathrm{factor}}>0$ is a parameter that sets the overall restoration speed.
This monotone, convex schedule grows slowly at early iterations and accelerates as $t\!\to\!T$, yielding a gradual center\mbox{-}to\mbox{-}edge restoration of macro density.
For $T\!=\!300$, the evolution of the macro density is shown in Figure~\ref{fig: exp density}.

As also shown in Figure~\ref{fig: exp density}, the charge distribution defined by Eqs.~\eqref{eq: exp_sigma} exhibits relatively sharp density changes near macro edges in late iterations. To mitigate these edge transients, we introduce two alternative \emph{charge-restoration} models, which are linear-like and sigmoid-like. They produce smoother boundary evolution.

\noindent\textbf{Linear-like model.}
\begin{equation}
\label{eq: linear_k}
\rho_{k}(\Delta x_{i,m},\Delta y_{i,m})
\;=\;
-\,k\,\sqrt{2(\dfrac{\Delta x_{i,m}^{2}}{w_m^{2}}+\dfrac{\Delta y_{i,m}^{2}}{h_m^{2}})} + 1 .
\end{equation}

\noindent\textbf{Sigmoid-like model.}
\begin{equation}\label{eq:sigmoid_k}
  \rho_{k}(\Delta x_{i,m},\Delta y_{i,m})
  \;=\;
  \frac{2}{1+\exp\!\left(
    k\,
      \sqrt{2(\dfrac{\Delta x_{i,m}^{2}}{w_m^{2}}+\dfrac{\Delta y_{i,m}^{2}}{h_m^{2}})}
  \right)} .
\end{equation}

Both models share the same iteration-dependent $k = k_{\mathrm{factor}}/{\tan\!\left(\dfrac{\pi}{2}\cdot\dfrac{t}{T}\right)}$, 
where $k_{\mathrm{factor}}>0$ is a parameter that sets the overall restoration speed.

In summary, we construct four iteration\mbox{-}dependent macro charge\mbox{-}density models $\rho_{t}(\Delta x_{i,m},\Delta y_{i,m})$. By design, these schedules promote a smooth evolution of bin densities while progressively restoring each macro’s footprint. We evaluate all four models in Section~IV\mbox{-}C.

Given $\rho_{t}$, the macro’s contribution to the density of bin $b$ at iteration $t$ is obtained by integrating over the bin region $B_b$:
\begin{equation}
  \Delta \rho^{(m)}_{b}(t)
  \;=\;
  \iint_{B_b}
  \rho_{t}(\Delta x_{i,m},\Delta y_{i,m})\,
  d(\Delta x)\, d(\Delta y).
\end{equation}
As illustrated in Figure~\ref{fig: Gaussion density}, the macro’s charge distribution progressively homogenizes over iterations. 
During early iterations (large \(\eta\)), the mass concentrates near the center, forming a narrow, high potential barrier that most standard cells can readily detour—effectively allowing them to traverse the macro region(Figure~\ref{fig: Gaussion density}(a)). 
In later iterations (small \(\eta\)), the density spreads to the full footprint, recovering the original macro. Only a small subset of standard cells with strong cross-macro connectivity can surmount this broader barrier, while most are guided to skirt the macro boundary(Figure~\ref{fig: Gaussion density}(d)).

\section{Autotuner}
\label{sec:autotuner}

\begin{figure}[!b]
  \centering
  \includegraphics[width=0.85\columnwidth]{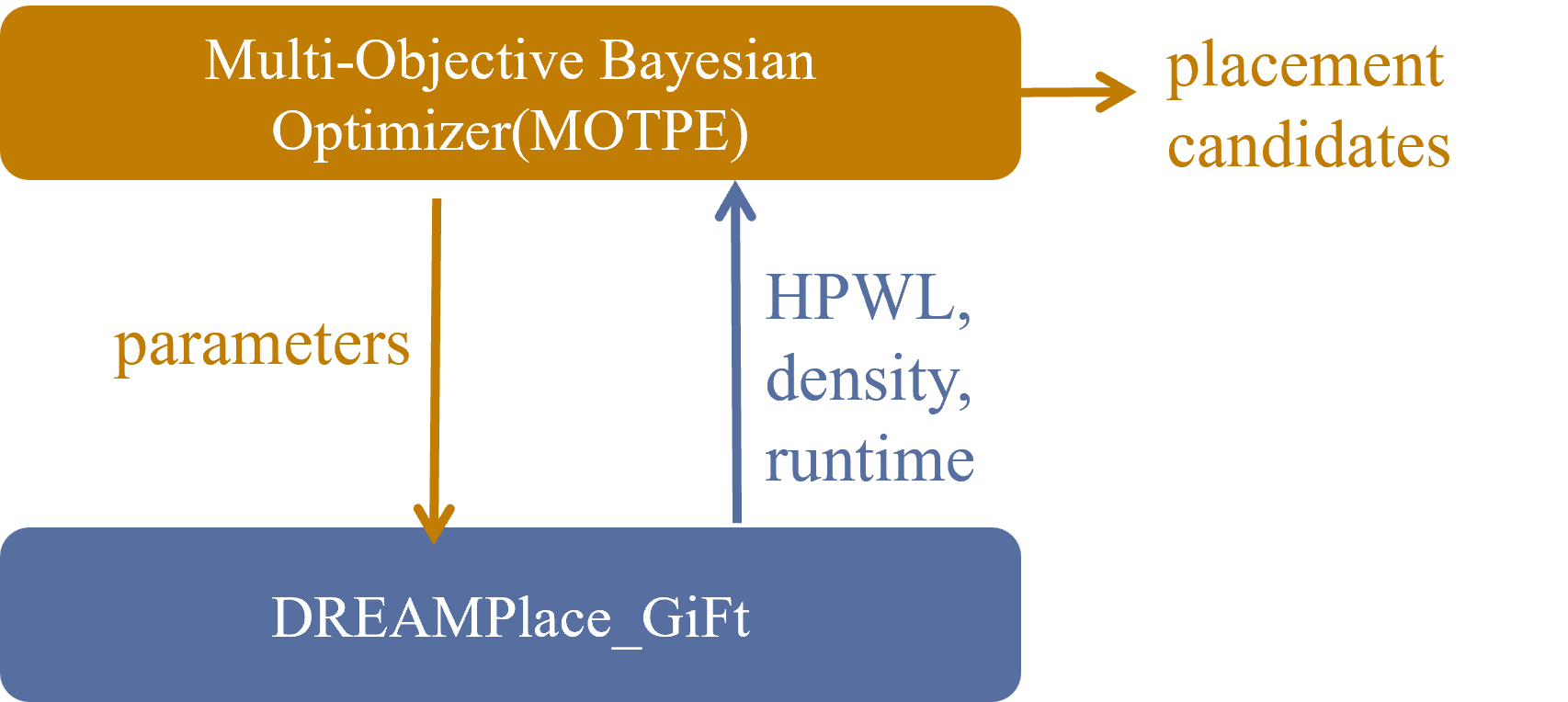}
  \caption{The three-stage optimization flow}
  \label{fig: tuner}
\end{figure}

Although a default parameter configuration is provided, design\mbox{-}specific tuning on each benchmark with our tuner can identify superior settings and yield improved results.
In this section, we propose a three-stage optimization methodology (Figure~\ref{fig: tuner}), built upon the AutoDMP~\cite{Agnesina2023AutoDMPAD} framework, to improve the quality of mixed-size global placement.
This approach integrates the multi-objective Bayesian optimization method to efficiently explore the high-dimensional design space and generate a diverse set of Pareto-optimal solutions that capture trade-offs among key performance metrics.

\begin{table}[t]
\centering
\caption{Parameter space for different stages}
\label{tab:placement_params}
\begin{tabularx}{\linewidth}{@{}>{\raggedright\arraybackslash}p{0.33\linewidth} X@{}}
\toprule
\textbf{Parameter} & \textbf{Explanation} \\
\midrule
\multicolumn{2}{c}{\textit{GSP\mbox{-}based initialization}} \\
\midrule
low\_filter\_sigma  & Self\mbox{-}loop coefficient for the low\mbox{-}band filter. \\
mid\_filter\_sigma  & Self\mbox{-}loop coefficient for the mid\mbox{-}band filter. \\
high\_filter\_sigma & Self\mbox{-}loop coefficient for the high\mbox{-}band filter. \\
low\_filter\_k      & Exponent $k$ in the eigenvalue response for the low\mbox{-}band filter. \\
mid\_filter\_k      & Exponent $k$ in the eigenvalue response for the mid\mbox{-}band filter. \\
high\_filter\_k     & Exponent $k$ in the eigenvalue response for the high\mbox{-}band filter. \\
low\_filter\_effect & Mixing weight for the low\mbox{-}band filter in the low/mid/high blend. \\
mid\_filter\_effect & Mixing weight for the mid\mbox{-}band filter in the low/mid/high blend (the high\mbox{-}band weight is $1-\texttt{low\_filter\_effect}-\texttt{mid\_filter\_effect}$). \\
\midrule
\multicolumn{2}{c}{\textit{Area\mbox{-}hint refinement}} \\
\midrule
refine\_iteration  & Number of refinement iterations ($T_{\mathrm{hint}}$). \\
refine\_num\_bin\_xy & Number of refinement bins along $x$/$y$ ($\mathrm{numBin}_x$, $\mathrm{numBin}_y$). \\
detection\_ratio   & Detection\mbox{-}window size ratio (fraction of $(\mathrm{numBin}_x,\mathrm{numBin}_y)$). \\
bin\_capacity      & Target per\mbox{-}bin density $\mathcal{C}$. \\
\midrule
\multicolumn{2}{c}{\textit{Macro\mbox{-}schedule placement}} \\
\midrule
schedule\_iteration & Number of scheduling iterations to restore macro footprints ($T$). \\
sigma\_factor       & Charge\mbox{-}restoration speed factor ($\sigma_{\mathrm{factor}}$). \\
\midrule
\multicolumn{2}{c}{\textit{DREAMPlace}} \\
\midrule
density\_weight     & Initial weight of the density term. \\
gamma               & Base coefficient for LSE/WA wirelength; scaled relative to bin size. \\
GP\_learning\_rate  & Initial learning rate for gradient\mbox{-}based global placement. \\
GP\_wirelength      & Wirelength model selection (LSE or WA). \\
RePlAce\_ref\_hpwl  & Reference HPWL used by RePlAce to adapt density weight. \\
RePlAce\_LOWER\_PCOF & Lower ratio threshold used by RePlAce when updating density weight. \\
RePlAce\_UPPER\_PCOF & Upper ratio threshold used by RePlAce when updating density weight. \\
\bottomrule
\end{tabularx}
\vspace{-2em}
\end{table}

\noindent
\textbf{Evaluation metrics}.
Our optimization problem is formulated as a multi-objective search, aiming to minimize three primary evaluation metrics:
We use HPWL measured after detailed placement as a assessment for wirelength.
We quantify the density by measuring the density overflow.
The computational cost of the placement is measured by the total runtime, which we aim to minimize for improved efficiency.


\noindent
\textbf{Search space}.
The search space includes the parameters in Table~\ref{tab:placement_params} regulating GSP-based initialization, area-hint refinement, and macro-schedule placement, as well as DREAMPlace’s built-in hyperparameters.

\noindent
\textbf{The three-stage optimization flow}. The proposed three-stage optimization process mainly includes the following three steps:
\begin{itemize}[noitemsep,topsep=0pt,leftmargin=*]

\item{Initialization:} The optimization is initialized from a known high-quality baseline configuration rather than a random starting point. 
This warm-start strategy reduces search overhead and accelerates convergence by directing the optimizer toward a promising region of the design space.

\item{Muliti-objective parameter space exploration:} The core search employs the Multi-Objective Tree-structured Parzen Estimator (MOTPE)~\cite{10.1145/3377930.3389817}, which models parameter–objective relations via two probability density functions, $L(x)$ (non-dominated solutions) and $G(x)$ (others). 
By maximizing $L(x)/G(x)$, MOTPE balances exploration and exploitation, efficiently discovering diverse Pareto-optimal solutions across HPWL, density and runtime.

\item{Solution distillation:} The Pareto front is post-processed using k-means clustering in the three-dimensional objective space (HPWL, density, runtime). 
A representative solution from each cluster is retained.
This step transforms a large Pareto front into a concise set of actionable candidates that capture key objective trade-offs.

\end{itemize}


\section{Experimental Validation}
\label{sec:exp}

Our framework is implemented in \texttt{C++} and \texttt{Python} on top of DREAMPlace~\cite{dramplace-github}.
All experiments are conducted on a server equipped with an Intel Xeon Gold~5320 CPU and five NVIDIA A800 (80\,GB, PCIe) GPUs.
We evaluate on the ISPD2005 contest suite and on two real-world designs (Ariane~\cite{ariane} and BP~Quad~\cite{bp_quad}) under both NanGate45 (NG45) and ASAP7 technology nodes.
For Ariane and BP~Quad, the macro placements are first generated by Hier-RTLMP~\cite{Hier-RTLMP}.
Benchmark characteristics are summarized in Table~\ref{tab:benchmark_characteristics}.
In Section~\ref{subsec:main-results}, we present the main experimental results. 
In Section~\ref{subsec: ablation study}, we conduct two ablation studies to quantify the roles of \emph{Area\mbox{-}Hint Refinement} and \emph{Macro\mbox{-}Schedule Placement}, and to compare the four macro\mbox{-}schedule models. 
In Section~\ref{subsec: sensitivity}, we assess the robustness of our framework under different random seeds and target\_density settings. 
Finally, in Section~\ref{subsec: finetuning}, we perform per\mbox{-}design tuning to further reduce HPWL.


\begin{table}[hbtp]
  \centering
  \scriptsize                           
  \setlength{\tabcolsep}{3pt}           
  \caption{Benchmark characteristics. ``Movable'' is the number of movable standard cells, while ``Fixed'' is the number of I/O pins and fixed macros.}
  \label{tab:benchmark_characteristics}
  \begin{tabular*}{\linewidth}{@{\extracolsep{\fill}}lrrrr}
    \toprule
    \textbf{Design} & \textbf{Cells} & \textbf{Nets} & \textbf{Movable} & \textbf{Fixed} \\
    \midrule
    adaptec1  & 211{,}990  & 221{,}142  & 211{,}447  & 543   \\
    adaptec2  & 255{,}589  & 266{,}009  & 255{,}023  & 566   \\
    adaptec3  & 452{,}373  & 466{,}758  & 451{,}650  & 723   \\
    adaptec4  & 497{,}374  & 515{,}951  & 496{,}045  & 1{,}329 \\
    bigblue1  & 278{,}724  & 284{,}479  & 278{,}164  & 560   \\
    bigblue2  & 580{,}950  & 577{,}235  & 557{,}866  & 23{,}084 \\
    bigblue3  & 1{,}098{,}110 & 1{,}123{,}170 & 1{,}096{,}812 & 1{,}298 \\
    bigblue4  & 2{,}185{,}523 & 2{,}229{,}886 & 2{,}177{,}353 & 8{,}170 \\
    \noalign{\hrule height 1.1pt}
    Ariane (NG45)           & 97{,}801    & 105{,}925   & 97{,}306    & 495   \\
    BP Quad (NG45)          & 715{,}573   & 768{,}634   & 715{,}438   & 135   \\
    Ariane (ASAP7)          & 98{,}341    & 99{,}709    & 97{,}846    & 495   \\
    BP Quad (ASAP7)         & 804{,}223   & 818{,}251   & 804{,}088   & 135   \\
    \bottomrule
  \end{tabular*}
\end{table}


\subsection{Main Results}
\label{subsec:main-results}

We evaluate DREAMPlace~\cite{gu2020dreamplace}, GiFt--DREAMPlace~\cite{liu2024power}, and our framework on the aforementioned benchmarks, reporting HPWL after detailed placement together with the number of global\mbox{-}placement (GP) iterations and the corresponding runtime. Results are summarized in Table~\ref{tab:results}. 

Previous work by Liu~\emph{et~al.}~\cite{liu2024power} reports that GiFt--DREAMPlace matches DREAMPlace on \emph{standard\mbox{-}cell\mbox{-}only} designs and achieves this with fewer GP iterations and shorter GP runtime. However, on our fourteen \emph{macro\mbox{-}heavy} benchmarks, we observe a noticeable HPWL degradation for GiFt.
In contrast, our framework achieves lower HPWL than DREAMPlace on these macro\mbox{-}heavy cases within a comparable runtime budget.

We also intended to compare against the initialization approach of Cheng~\emph{et~al.}~\cite{chen2023placement}. We contacted the authors and they are working on the release. But as of submission, the release is not finished yet. Based on the experiment result reported in their paper, their method achieves an average 2.2\% wirelength reduction on the ISPD contest suite, which exceed our 1\% improvement. However, their initialization stage is reported to be time\mbox{-}consuming (sometimes longer than placement itself)~\cite{chen2022placement}. By contrast, our initialization is much less time-cosuming, the exact initialization run time can be seen in Table~\ref{tab:init-time}.

\begin{table}[hbtp]
\centering
\caption{Initialization time per benchmark.}
\label{tab:init-time}
\scriptsize
\setlength{\tabcolsep}{4pt}
\renewcommand{\arraystretch}{1.08}
\begin{tabularx}{\linewidth}{l r l r}
\toprule
\textbf{Benchmark} & \textbf{Init. Time(s)} & \textbf{Benchmark} & \textbf{Init. Time(s)} \\
\midrule
adaptec1 & 0.450 & bigblue1 & 0.507 \\
adaptec2 & 3.609   & bigblue2 & 1.147 \\
adaptec3 & 1.266 & bigblue3 & 12.158 \\
adaptec4 & 1.317   & bigblue4 & 19.604 \\
Ariane (NG45) & 1.183 & BP Quad (NG45) & 3.035 \\
Ariane (ASAP7) & 2.391 & BP Quad (ASAP7) & 2.460 \\
\bottomrule
\end{tabularx}
\end{table}

\begin{table*}[hbtp]
\centering
\caption{Cumulative post-detailed-placement HPWL, GP runtime, and GP iterations.}
\label{tab:results}
\scriptsize
\setlength{\tabcolsep}{2.5pt}
\renewcommand{\arraystretch}{1.12}
\begin{tabular*}{\textwidth}{@{\extracolsep{\fill}}lccc ccc ccc@{}}
\toprule
\multirow{2}{*}{\textbf{Design}} &
\multicolumn{3}{c}{\textbf{\shortstack{DREAMPlace\\(rand.)}}} &
\multicolumn{3}{c}{\textbf{\shortstack{GiFt--\\DREAMPlace}}} &
\multicolumn{3}{c}{\textbf{\shortstack{Our\\framework}}} \\
\cmidrule(lr){2-4}\cmidrule(lr){5-7}\cmidrule(l){8-10}
& \textbf{HPWL} & \textbf{Time (s)} & \textbf{Iter.}
& \textbf{HPWL} & \textbf{Time (s)} & \textbf{Iter.}
& \textbf{HPWL} & \textbf{Time (s)} & \textbf{Iter.} \\
\midrule
adaptec1   & 72{,}767{,}740  & 7.267  & 618 & 73{,}079{,}130(-0.4\%)  & 10.668 & 439 & \textcolor{blue}{72{,}596{,}350\,(0.24\%)}  & 12.567 & 682 \\
adaptec2   & 81{,}891{,}890  & 8.747  & 643 & 81{,}485{,}180(0.5\%)   & 12.839 & 444 &  \textcolor{blue}{80{,}926{,}890\,(1.20\%)}  & 26.155 & 607 \\
adaptec3   & 192{,}870{,}300 & 12.128 & 680 & 194{,}118{,}500(-0.6\%) & 19.944 & 457 &  \textcolor{blue}{190{,}355{,}700\,(1.32\%)} & 21.921 & 1001 \\
adaptec4   & 173{,}542{,}700 & 12.134 & 718 & 177{,}011{,}660(-2\%)   & 18.978 & 480 &  \textcolor{blue}{171{,}749{,}300\,(1.04\%)} & 17.513 & 858 \\
bigblue1   & 89{,}276{,}640  & 7.872  & 680 & 89{,}046{,}700(-0.2\%)  & 14.400 & 476 & 89{,}598{,}570\,(-0.30\%)  & 12.097 & 646 \\
bigblue2   & 136{,}975{,}700 & 17.022 & 665 & 140{,}039{,}000(-2.2\%) & 26.409 & 491 &  \textcolor{blue}{136{,}569{,}700\,(0.30\%)} & 27.194 & 816 \\
bigblue3   & 303{,}963{,}100 & 21.071 & 816 & 314{,}034{,}500(-3.3\%) & 39.015 & 600 &  \textcolor{blue}{297{,}366{,}600\,(2.22\%)} & 51.636 & 892 \\
bigblue4   & 742{,}645{,}400 & 35.232 & 848 & 768{,}316{,}300(-3.3\%) & 91.985 & 600 &  \textcolor{blue}{739{,}276{,}200\,(0.46\%)} & 93.706 & 885 \\
\noalign{\hrule height 1.2pt}
Ariane (NG45)   & 13{,}625{,}530 & 5.228 & 597 & 13{,}583{,}270(0.3\%) & 7.189  & 429 &  \textcolor{blue}{13{,}540{,}130\,(0.63\%)} & 8.434  & 738 \\
BP Quad (NG45)  & 95{,}886{,}810  & 20.197  & 706 & 96{,}156{,}230(-0.3\%)  & 22.707 & 472 &  \textcolor{blue}{94{,}647{,}100\,(1.31\%)} & 33.441 & 935 \\
Ariane (ASAP7)  & 9{,}884{,}404   & 12.316 & 541 & 9{,}832{,}253(0.5\%)  & 11.886 & 430 &  \textcolor{blue}{9{,}680{,}402\,(2.11\%)}  & 12.405 & 410 \\   
BP Quad (ASAP7) & 85{,}488{,}340 & 33.703 & 697 & 88{,}297{,}780(-3.2\%)  & 38.558 & 475 & \textcolor{blue}{84{,}776{,}460\,(0.84\%)} & 40.611 & 872 \\
\bottomrule
\end{tabular*}
\end{table*}

From Table~\ref{tab:results} and Table~\ref{tab:init-time}, we draw the following conclusions:
\begin{itemize}[noitemsep,topsep=0pt,leftmargin=*]
  \item \textbf{Post-placement wirelength.} On \emph{macro\mbox{-}heavy} benchmarks, GiFt--DREAMPlace shows noticeable HPWL degradation. In contrast, our framework achieves lower post\mbox{-}detailed\mbox{-}placement HPWL than DREAMPlace on all cases \emph{except} \textit{bigblue1} (where it is marginally worse by $0.30\%$); the best improvement is $2.22\%$, with an average improvement of about $1\%$.
  \item \textbf{Placement iterations and total runtime.} Our framework increases GP iterations by about $20\%$ on average. However, consistent with the observations for GiFt--DREAMPlace, a reduction in iterations on macro\mbox{-}heavy designs does not necessarily translate into shorter runtime. Overall, our runtime remains within an acceptable, comparable budget. For the largest case, \emph{bigblue4} (approximately $2.2$M cells), the GP takes about $90$\,s. 
  \item \textbf{Initialization runtime.} The initialization stage of our method is lightweight: for designs with fewer than one million gates, initialization completes within $10$\,s. Compared with the \emph{reported} initialization runtime in Chen~\emph{et~al.}~\cite{chen2022placement}\, our initializer is nearly two orders of magnitude ($\sim\!100\times$) faster. While our hardware differs from theirs and we cannot reproduce their exact setup, this comparison still indicates that our initializer has substantially lower computational complexity than area-aware initialization. Its cost is far smaller than the placement runtime and therefore does not become a time\mbox{-}wall bottleneck in the flow.
\end{itemize}

\subsection{Ablation Study}
\label{subsec: ablation study}


\begin{figure}[hbtp]
  \centering
  \subfloat[]{
    \includegraphics[width=\linewidth]{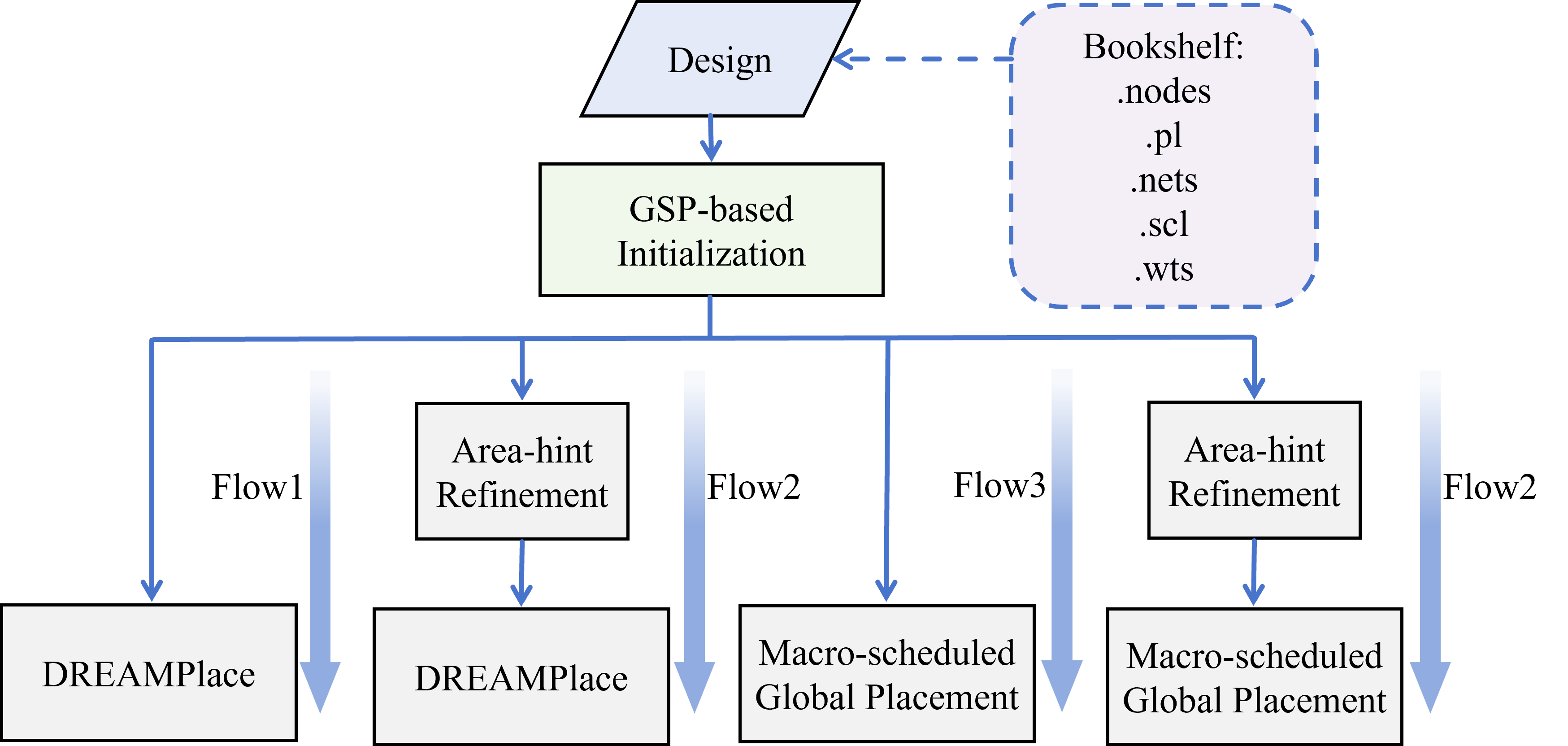}
  }

  \par\medskip

  \subfloat[]{
    \includegraphics[width=\linewidth]{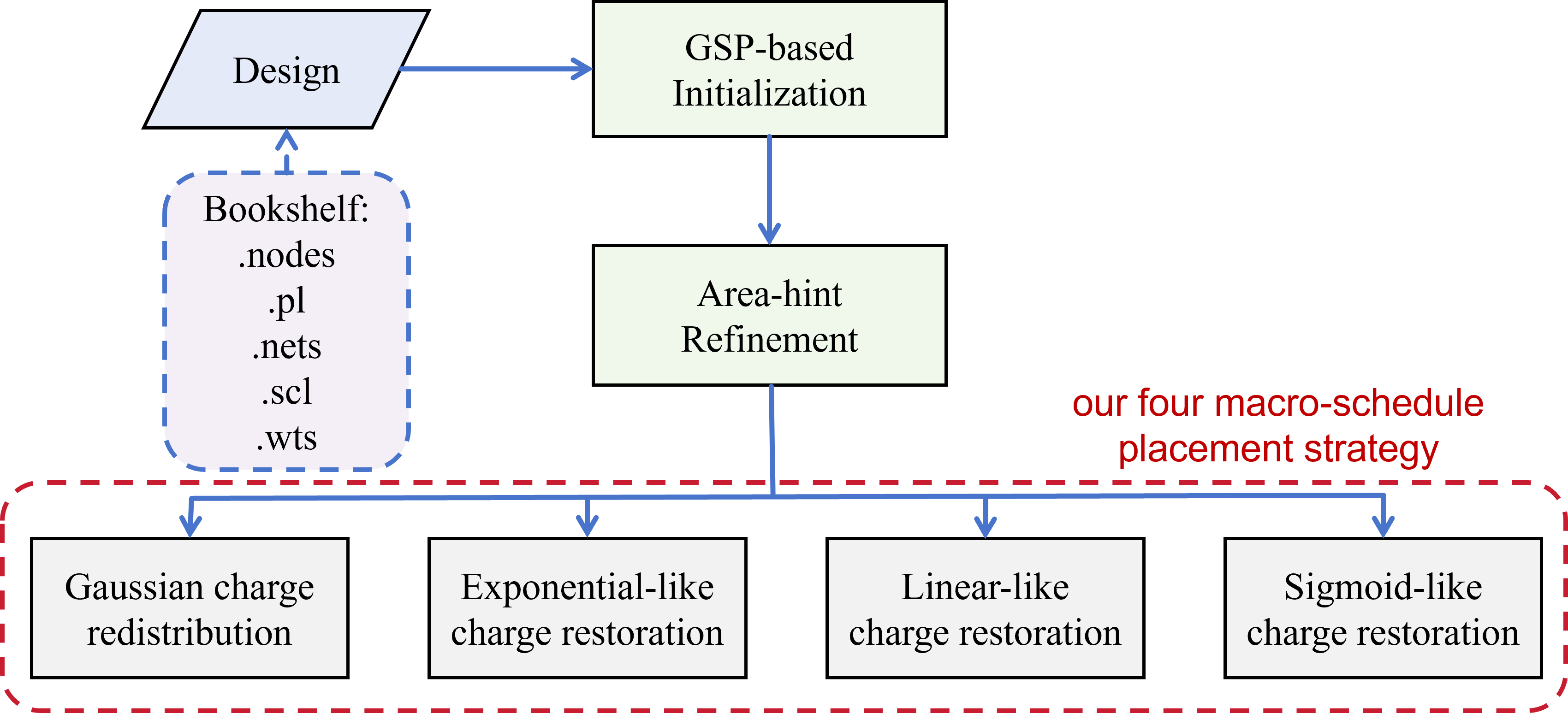}
  }
  
  \caption{The evaluation flow for ablation studies}
  \label{fig: abaltion study}
\end{figure}

\definecolor{BetterBG}{HTML}{E6F4EA} 
\definecolor{WorseBG}{HTML}{FDECEA}  
\newcommand{\better}[1]{\cellcolor{BetterBG}{#1}}
\newcommand{\worse}[1]{\cellcolor{WorseBG}{#1}}

\begin{table}[hbtp]
\centering
\caption{Post-detailed-placement HPWL for ablation settings.}
\label{tab:ablation-study}
\scriptsize
\setlength{\tabcolsep}{6pt}
\renewcommand{\arraystretch}{1.12}
\begin{tabularx}{\linewidth}{lXX}
\toprule
\textbf{Design} & \textbf{Only Area-hint refine.} & \textbf{Only Macro-sched. place.} \\
\midrule
adaptec1  & \better{72435390} & \worse{74979070} \\
adaptec2  & \better{81039480} & \better{81042960} \\
adaptec3  & \better{190028200} & \better{190977200} \\
adaptec4  & \better{172026100} & \worse{176493300} \\
bigblue1  & \better{89152700} & \better{89249870} \\
bigblue2  & \worse{139066300} & \worse{137546200} \\
bigblue3  & \worse{305731600} & \worse{304607900} \\
bigblue4  & \worse{748730200} & \worse{1757296000} \\
Ariane (NG45)  & \worse{13842280} & \better{13590720} \\ 
BP Quad (NG45) & \better{94284160} & \better{94561150} \\
Ariane (ASAP7) & \better{9696391}  & \better{9740737} \\
BP Quad (ASAP7) & \better{84820800} & \worse{87953120} \\
\bottomrule
\end{tabularx}
\end{table}

In this subsection, we conduct two ablation studies.
We quantify the individual contributions of \emph{Area\mbox{-}Hint Refinement} and \emph{Macro\mbox{-}Schedule Placement} to HPWL. The evaluation pipeline is shown in Figure~\ref{fig: abaltion study}(a). We report post\mbox{-}detailed\mbox{-}placement HPWL for four settings: \emph{(Flow1)} GSP\mbox{-}based Initialization (GiFt) directly followed by DREAMPlace; \emph{(Flow2)} DREAMPlace with Area\mbox{-}Hint Refinement only; \emph{(Flow3)} DREAMPlace with Macro\mbox{-}Schedule Placement only; and \emph{(Flow4)} the full proposed framework. Results are summarized in Table~\ref{tab:ablation-study}, where green cells denote lower (better) post\mbox{-}detailed\mbox{-}placement HPWL than DREAMPlace, and red cells denote higher (worse) HPWL. Note that \emph{Flow1} and \emph{Flow4} duplicate the corresponding entries in Table~\ref{tab:results}; to avoid redundancy, they are omitted from Table~\ref{tab:ablation-study}.

Building on this analysis, we then compare the four macro\mbox{-}scheduling models introduced in Section~III\mbox{-}C (Gaussian charge\mbox{-}redistribution, exponential\mbox{-}like charge\mbox{-}restoration, linear\mbox{-}like charge\mbox{-}restoration, and sigmoid\mbox{-}like charge\mbox{-}restoration) following the pipeline in Figure~\ref{fig: abaltion study}(b). The comparison is reported in Table~\ref{tab: macro schedule strategy}.


\begin{figure}[htbp]
  \centering
  \includegraphics[width=\columnwidth]{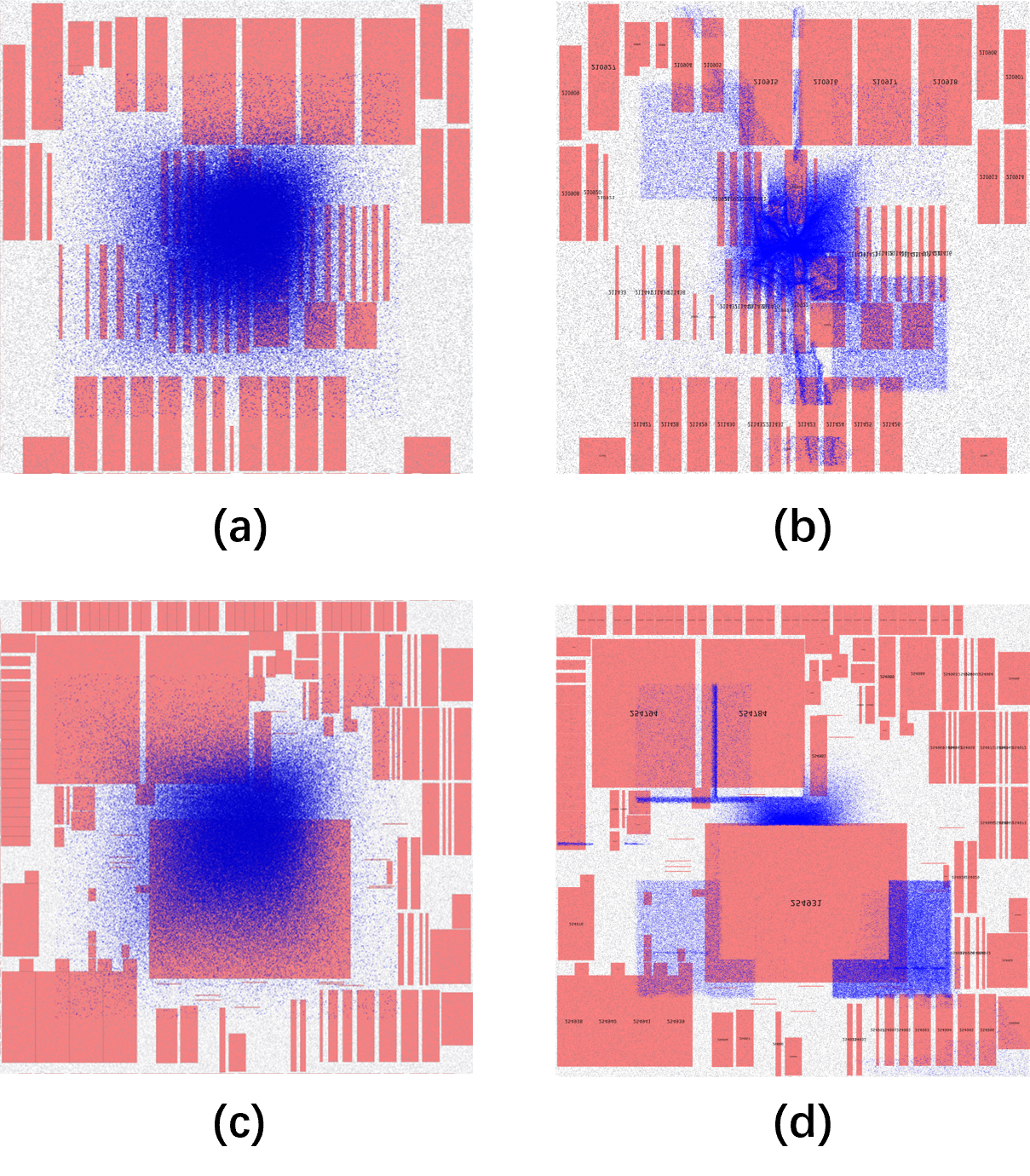}
  \caption{Initialization comparison between GiFt and GiFt + Area-Hint Refinement.  Red denotes fixed macros and I/O pins; blue denotes movable standard cells. 
(a) \textit{adaptec1} (GiFt); (b) \textit{adaptec1} (GiFt + Area-Hint Refinement); 
(c) \textit{adaptec2} (GiFt); (d) \textit{adaptec2} (GiFt + Area-Hint Refinement).
}
  \label{fig:Area-hint Refinement}
\end{figure}

\noindent
\textbf{Area\mbox{-}Hint Refinement}.
Comparing \emph{Flow1} and \emph{Flow2}, we observe that inserting the Area\mbox{-}Hint Refinement stage between the GSP\mbox{-}based initialization and DREAMPlace yields a clear reduction in post\mbox{-}detailed\mbox{-}placement HPWL. This indicates that the refinement successfully injects useful heuristic area information into the graph signals generated by GSP-Based Initialization. 
Figure~\ref{fig:Area-hint Refinement} contrasts GiFt with GiFt\,+\,Area\mbox{-}Hint Refinement: GiFt produces nearly identical, center\mbox{-}concentrated placements under different macro distributions, whereas Area\mbox{-}Hint encourages movable cells to diffuse toward macro\mbox{-}free regions. 
The visible boundaries in the plots arise from the bin partition used to construct the hints. This is expected and acceptable, since initialization only needs to coarsely determine the bin\mbox{-}level locations of movable cells.

Howerver, as shown in Table~\ref{tab:ablation-study}, \emph{Flow2} still underperforms plain DREAMPlace on some designs. We attribute this to a mismatch between our initialization and the subsequent global placer-the gap between initialization and global placement. To bridge this gap, we introduce \emph{macro\mbox{-}schedule placement}.

\noindent
\textbf{Macro-schedule Placement}. Comparing \emph{Flow1} and \emph{Flow3}, we observe that adding the macro\mbox{-}schedule placement stage often yields a noticeable reduction in post\mbox{-}detailed\mbox{-}placement HPWL on some designs. We attribute this to the very early iterations of macro\mbox{-}scheduling, where the smoothed macro model can still exhibit numerous narrow spikes. Because the GSP\mbox{-}based initialization is not area\mbox{-}aware, some cells may be initialized on top of these spikes, causing undesirable perturbations. The \emph{Area\mbox{-}Hint Refinement} prevents this failure mode by steering movable cells away from macro regions at the bin level, thereby providing a starting point that is more compatible with macro\mbox{-}scheduling. 

Overall, the first ablation demonstrates that both components contribute positively; used together, they bridge the representational gap between the GSP\mbox{-}based point model in initialization and the area\mbox{-}aware analytical model in global placement.

\definecolor{DeepGreen}{HTML}{0B6B3A} 
\newcommand{\gframe}[1]{%
  \tikz[baseline=(c.base)]%
    \node[draw=DeepGreen, line width=0.8pt, rounded corners=2pt,
          inner sep=1pt, outer sep=0pt, fill=none] (c) {\strut #1};}

\begin{table*}[hbtp]
\centering
\caption{Post\mbox{-}detailed\mbox{-}placement HPWL for macro\mbox{-}scheduling models.}
\label{tab: macro schedule strategy}
\scriptsize
\setlength{\tabcolsep}{6pt}
\renewcommand{\arraystretch}{1.12}
\begin{tabular*}{\textwidth}{@{\extracolsep{\fill}}lrrrr@{}}
\toprule
\textbf{Design} &
\textbf{Gaussian redistribution} &
\textbf{Exponential\mbox{-}like restoration} &
\textbf{Linear\mbox{-}like restoration} &
\textbf{Sigmoid\mbox{-}like restoration} \\
\midrule
adaptec1  & \(7.30754\times 10^{7}\)\,(-0.42\%) & \textcolor{blue}{\(7.25963\times 10^{7}\)\,(0.24\%)}  & \(7.32467\times 10^{7}\)\,(-0.65\%) & \(7.26280\times 10^{7}\)\,(0.19\%) \\
adaptec2  & \(8.15964\times 10^{7}\)\,(0.31\%)  & \textcolor{blue}{\(8.09268\times 10^{7}\)\,(1.20\%)}  & \(8.23302\times 10^{7}\)\,(-0.53\%) & \(8.09526\times 10^{7}\)\,(1.16\%) \\
adaptec3  & \(1.94608\times 10^{8}\)\,(-0.89\%) & \textcolor{blue}{\(1.90355\times 10^{8}\)\,(1.32\%)} & \(1.94252\times 10^{8}\)\,(-0.71\%) & \(1.90503\times 10^{8}\)\,(1.24\%) \\
adaptec4  & \(1.75795\times 10^{8}\)\,(-1.28\%) & \textcolor{blue}{\(1.71749\times 10^{8}\)\,(1.04\%)} & \(1.76494\times 10^{8}\)\,(-1.67\%) & \(1.72555\times 10^{8}\)\,(0.57\%) \\
bigblue1  & \textcolor{blue}{\(8.91383\times 10^{7}\)\,(0.15\%)}  & {\(8.95985\times 10^{7}\)\,(-0.30\%)} & \(8.92200\times 10^{7}\)\,(0.06\%)  & \(8.96880\times 10^{7}\)\,(-0.40\%) \\
bigblue2  & \(1.39119\times 10^{8}\)\,(-1.54\%) & \textcolor{blue}{\(1.36569\times 10^{8}\)\,(0.30\%)} & \(1.38630\times 10^{8}\)\,(-1.19\%) & \(1.36618\times 10^{8}\)\,(0.26\%) \\
bigblue3  & \(3.09728\times 10^{8}\)\,(-1.86\%) & \textcolor{blue}{\(2.97366\times 10^{8}\)\,(2.22\%)} & \(3.09768\times 10^{8}\)\,(-1.87\%) & \(2.97689\times 10^{8}\)\,(2.11\%) \\
bigblue4  & \(7.57456\times 10^{8}\)\,(-1.96\%) & \(7.39276\times 10^{8}\)\,(0.46\%) & \(7.60459\times 10^{8}\)\,(-2.34\%) & \textcolor{blue}{\(7.39272\times 10^{8}\)\,(0.46\%)} \\
Ariane (NG45)  & \(1.38233\times 10^{7}\)\,(-1.43\%) & \textcolor{blue}{\(1.35401\times 10^{7}\)\,(0.63\%)} & \(1.42327\times 10^{7}\)\,(-4.27\%) & \(1.35734\times 10^{7}\)\,(0.38\%) \\
BP Quad (NG45) & \(1.00479\times 10^{8}\)\,(-4.57\%) & \(9.46471\times 10^{7}\)\,(1.31\%) & \(1.01407\times 10^{8}\)\,(-5.44\%) & \textcolor{blue}{\(9.46305\times 10^{7}\)\,(1.33\%)} \\
Ariane (ASAP7) & \(9.88812\times 10^{6}\)\,(-0.38\%) & \textcolor{blue}{\(9.68040\times 10^{6}\)\,(2.11\%)}  & \(1.03418\times 10^{7}\)\,(-4.42\%) & \(9.72576\times 10^{6}\)\,(1.63\%) \\
BP Quad (ASAP7) & \(8.67922\times 10^{7}\)\,(-1.50\%) & \textcolor{blue}{\(8.47764\times 10^{7}\)\,(0.84\%)} & \(9.07764\times 10^{7}\)\,(-5.83\%) & \(8.54539\times 10^{7}\)\,(0.04\%) \\
\bottomrule
\end{tabular*}
\end{table*}

\noindent
\textbf{Macro-schedule Model Comparison}. Table~\ref{tab: macro schedule strategy} reports results for the four macro\mbox{-}scheduling models. Values in parentheses denote HPWL change relative to DREAMPlace. For most designs, all four schedules yield shorter HPWL than the baseline that directly connects the GSP\mbox{-}based initialization to the global placer. The per\mbox{-}design best HPWL is highlighted in blue. Overall, \emph{restoration\mbox{-}based} schedules outperform \emph{redistribution}, consistent with our analysis in Section~III\mbox{-}C that redistribution can induce excessively high potential barriers in early iterations. Within the restoration family, both \emph{exponential\mbox{-}like} and \emph{sigmoid\mbox{-}like} models reduce HPWL, with the \emph{exponential\mbox{-}like} model being superior on most designs. Accordingly, we adopt the exponential\mbox{-}like restoration as the default in our flow.


\begin{figure}[hbtp]
  \centering
  \subfloat[bigblue4]{%
    \includegraphics[width=\linewidth]{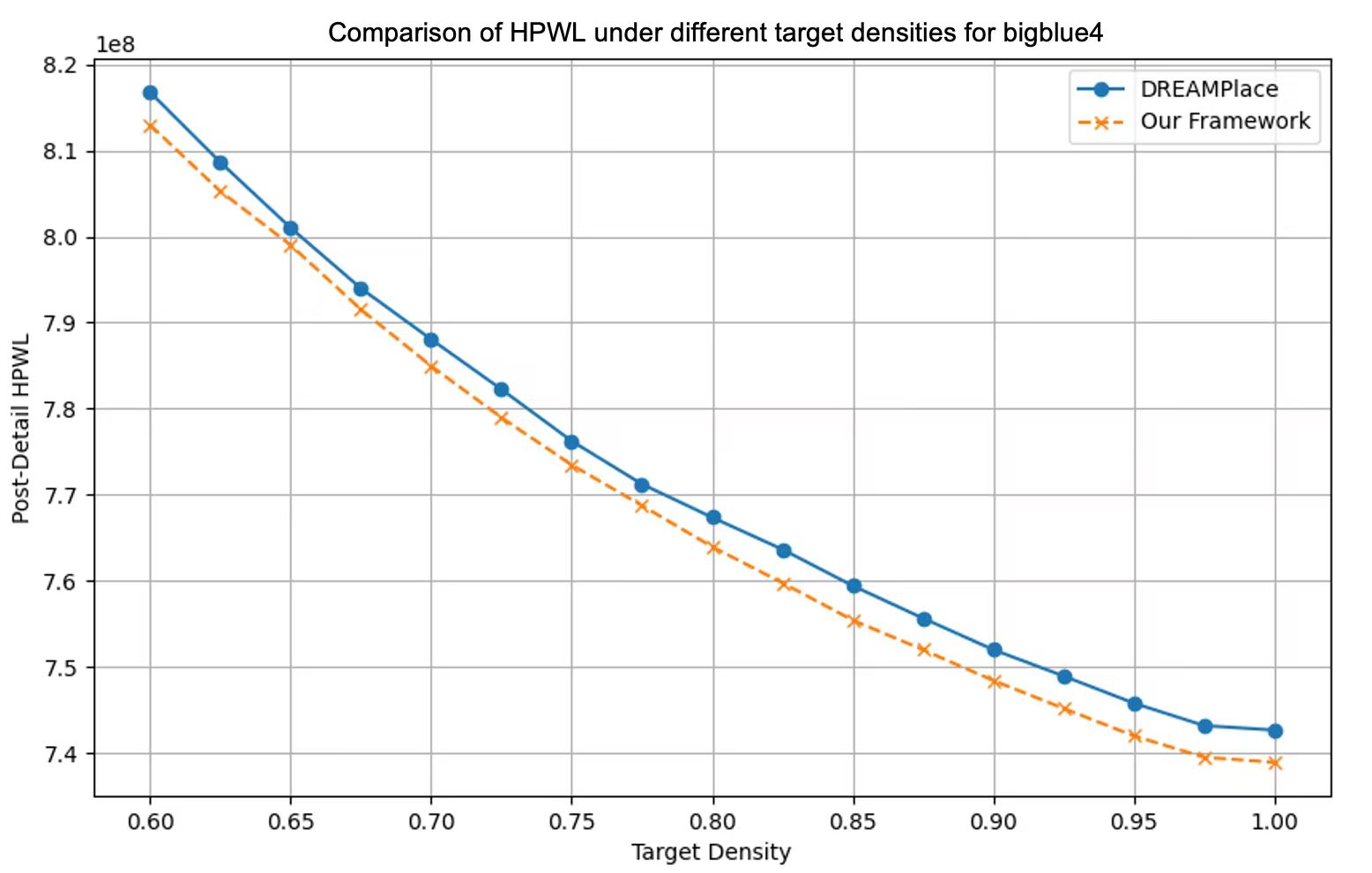}%
    \label{fig:td_bb4_hpwl}}%
  \vspace{-0.35\baselineskip}   
  \subfloat[BP Quad (NG45)]{%
    \includegraphics[width=\linewidth]{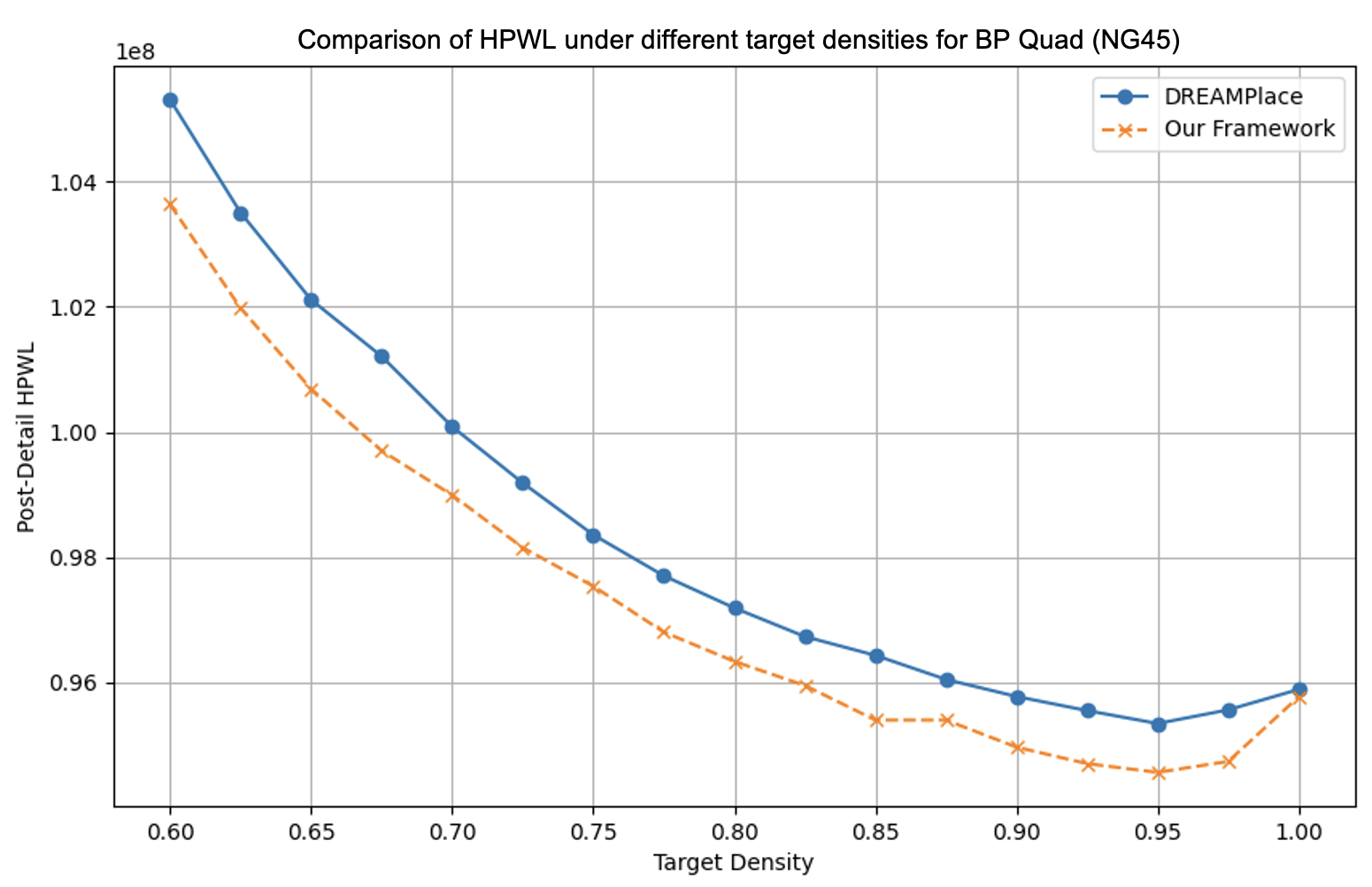}%
    \label{fig:td_bb4_runtime}}
  \caption{Comparison between our framework and DREAMPlace across target\_density settings for \textit{bigblue4} and \textit{BP Quad (NG45)}.}
  \label{fig:traget_density}
\end{figure}

\subsection{Sensitivity study}
\label{subsec: sensitivity}

\noindent
\textbf{Sensitivity to Target Density}.
Global placement is known to be sensitive to the \texttt{target\_density} parameter. 
We sweep \texttt{target\_density} from $0.60$ to $1.00$ in steps of $0.25$ and compare our framework with DREAMPlace. 
Although HPWL varies with the density budget, our framework attains shorter HPWL at nearly all settings. 
Figure~\ref{fig:traget_density} shows the post\mbox{-}detailed\mbox{-}placement HPWL for \textit{bigblue4} and \textit{BP Quad (NG45)}, illustrating the consistent advantage of our approach across densities.

\noindent
\textbf{Sensitivity to random seeds}.
As noted in Section~III\mbox{-}A, the initial graph signals in the GSP\mbox{-}based initialization are randomly sampled. Using the ISPD\mbox{2005} suite, we generate fifty random initial graph signals per benchmark and report the minimum, maximum, and mean post\mbox{-}detailed\mbox{-}placement HPWL across runs (Table~\ref{tab: random}). 
The observed variation is small, indicating that our framework is \emph{insensitive to random seeds} and consistently delivers stable, high\mbox{-}quality results.

\begin{table}[h]
\centering
\caption{Sensitivity to random seeds on ISPD2005: minimum, maximum, and average post\mbox{-}detailed\mbox{-}placement HPWL over 50 runs. The last column reports the relative range $(\max-\min)/\mathrm{avg}$.}
\label{tab: random}
\scriptsize
\setlength{\tabcolsep}{4pt}
\renewcommand{\arraystretch}{1.12}
\begin{tabularx}{\linewidth}{lrrrr}
\toprule
\textbf{Case} & \textbf{Min\_HPWL} & \textbf{Max\_HPWL} & \textbf{Avg\_HPWL} & \textbf{range/avg} \\
\midrule
adaptec1 & 72{,}579{,}940 & 72{,}628{,}720 & 72{,}603{,}798   & 0.000671866 \\
adaptec2 & 80{,}899{,}950 & 80{,}971{,}820 & 80{,}930{,}145.4 & 0.00088805 \\
adaptec3 & 190{,}019{,}200 & 190{,}751{,}200 & 190{,}346{,}092  & 0.003845627 \\
adaptec4 & 171{,}720{,}100 & 171{,}993{,}300 & 171{,}796{,}666  & 0.001590252 \\
bigblue1 & 89{,}578{,}100  & 89{,}621{,}620  & 89{,}597{,}611   & 0.000485727 \\
bigblue2 & 136{,}574{,}300 & 136{,}734{,}800 & 136{,}640{,}834  & 0.001174612 \\
bigblue3 & 297{,}296{,}900 & 297{,}507{,}600 & 297{,}407{,}820.5& 0.000708455 \\
bigblue4 & 739{,}085{,}600 & 739{,}546{,}500 & 739{,}326{,}114.3& 0.000623406 \\
\bottomrule
\end{tabularx}
\end{table}

\subsection{Finetuning}
\label{subsec: finetuning}
Leveraging the tuner described in Section~III\mbox{-}D, we perform design\mbox{-}specific hyperparameter tuning. Using two real\mbox{-}world designs across two technology nodes as case studies, we run 400 iterations of Bayesian optimization per design, adjusting the parameters listed in Table~\ref{tab:placement_params}. The results in Table~\ref{tab: tuner} show a reduction in post\mbox{-}detailed\mbox{-}placement HPWL for every case, with an average improvement of approximately \(0.15\%\). These findings indicate that analyzing design characteristics and selecting tailored parameters further enhance the effectiveness of our framework.

\begin{table}[hbtp]
\centering
\caption{Effect of Auto Tuner.}
\label{tab: tuner}
\scriptsize
\setlength{\tabcolsep}{6pt}
\renewcommand{\arraystretch}{1.12}
\begin{tabularx}{\linewidth}{l r r}
\toprule
\textbf{Design} & \textbf{Without tuner} & \textbf{With tuner} \\
\midrule
Ariane (NG45)    & 13,540,134 (0.63\%) & 13,513,482 (0.83\%) \\
BP Quad (NG45)   & 94,647,104 (1.31\%) & 94,554,496 (1.41\%) \\
Ariane (ASAP7)   & 9,680,402 (2.11\%)  & 9,668,405 (2.23\%) \\
BP Quad (ASAP7)  & 84,776,464 (0.84\%) & 84,541,568 (1.12\%) \\
\bottomrule
\end{tabularx}
\end{table}

\section{Conclusion and future work}
In this work, we introduced a lightweight co-optimization framework that narrows the gap between fast point-based initialization and area-aware analytical global placement. The approach combines two coordinated strategies, which are \textbf{area-hint refinement} and \textbf{macro-scheduled placement}. This smooth transition converts a topology-revealing point model into a physically consistent area-aware objective without incurring the heavy cost of fully area-aware initializers.
Across the ISPD2005 mixed-size suite and two real-world designs over two technology nodes (12 cases in total), the framework improves HPWL over DREAMPlace in 11 of 12 cases, with up to 2.2\% reduction, while keeping the runtime substantially lower than that of area-aware initializers.
This study suggests that full and exact area modeling during initialization is not strictly necessary. A smooth transition from a fast point-based model to the full area-aware objective yields strong convergence and high-quality solutions efficiently.

Future directions include the following.
\begin{enumerate}
  \item \textbf{Further speedups.} Although markedly faster than fully area-aware initializers, our framework still trails DREAMPlace’s point-based initializer. We will accelerate signed-graph signal filtering by \emph{sparsifying} the signed-graph Laplacian—pruning it to retain only the most salient negative edges that encode area information, thereby reducing the cost of preserving spectral structure. In addition, we will design a early-stop criterion tailored to macro-scheduled placement to terminate global placement (GP) iterations earlier without quality loss.

  \item \textbf{3D-IC extension.} Our current framework targets 2D ICs. To support 3D ICs, for example face-to-face (F2F) placement, the initializer must at a minimum encode hybrid-bonding constraints, tier-specific density targets, and cross-tier nets within the signed-graph.
\end{enumerate}



\medskip

\medskip

\begin{thebibliography}{99}

\bibitem{replace}
C.-K. Cheng, A. B. Kahng, I. Kang, and L. Wang, ``Replace: Advancing solution quality and routability validation in global placement'', 
\emph{IEEE Trans.CAD}, 38(9), 2018, pp. 1717--1730.

\bibitem{chen2022placement}
P. Chen, C.-K. Cheng, A. Chern, C. Holtz, A. Li, and Y. Wang, ``Placement initialization via a projected eigenvector algorithm: late breaking results'', 
\emph{Proc.DAC}, 2022, pp. 1398--1399.

\bibitem{chen2023placement}
P. Chen, C.-K. Cheng, A. Chern, C. Holtz, A. Li, and Y. Wang, ``Placement Initialization via Sequential Subspace Optimization with Sphere Constraints'', 
\emph{Proc. ISPD}, 2023, pp. 133--140.

\bibitem{chen2008ntuplace3}
T.-C. Chen, Z.-W. Jiang, T.-C. Hsu, H.-C. Chen, and Y.-W. Chang, ``NTUplace3: An analytical placer for large-scale mixed-size designs with preplaced blocks and density constraints'', 
\emph{IEEE Trans.CAD}, 27(7), 2008, pp. 1228--1240.

\bibitem{cheung2018robust}
G. Cheung, W.-T. Su, Y. Mao, and C.-W. Lin, ``Robust semisupervised graph classifier learning with negative edge weights'', 
\emph{TSIPN}, 4(4), 2018, pp. 712--726.

\bibitem{dong2016learning}
X. Dong, D. Thanou, P. Frossard, and P. Vandergheynst, ``Learning Laplacian matrix in smooth graph signal representations'', 
\emph{IEEE Trans. Signal Process.}, 64(23), 2016, pp. 6160--6173.

\bibitem{furutani2019graph}
S. Furutani, T. Shibahara, M. Akiyama, K. Hato, and M. Aida, ``Graph signal processing for directed graphs based on the hermitian laplacian'', 
\emph{Proc. ECML PKDD}, 2019, pp. 447--463.

\bibitem{gu2020dreamplace}
J. Gu, Z. Jiang, Y. Lin, and D. Z. Pan, ``DREAMPlace 3.0: Multi-electrostatics based robust VLSI placement with region constraints'', 
\emph{Proc. ICCAD}, 2020, pp. 1--9.

\bibitem{haynsworth1968inertia}
E. V. Haynsworth and A. M. Ostrowski, ``On the inertia of some classes of partitioned matrices'', 
\emph{Linear Algebra and its Applications}, 1(2), 1968, pp. 299--316.

\bibitem{hsu2011tsv}
M.-K. Hsu, Y.-W. Chang, and V. Balabanov, ``TSV-aware analytical placement for 3D IC designs'', 
\emph{Proc. DAC}, 2011, pp. 664--669.

\bibitem{liu2022graphplanner}
Y. Liu, Z. Ju, Z. Li, M. Dong, H. Zhou, J. Wang, F. Yang, X. Zeng, and L. Shang, ``Graphplanner: Floorplanning with graph neural network'', 
\emph{TODAES}, 28(2), 2022, pp. 1--24.

\bibitem{liu2023personalized}
J. Liu, D. Li, H. Gu, T. Lu, P. Zhang, L. Shang, and N. Gu, ``Personalized graph signal processing for collaborative filtering'', 
\emph{Proc. ACM Web Conference 2023}, 2023, pp. 1264--1272.

\bibitem{liu2024power}
Y. Liu, H. Zhou, J. Wang, F. Yang, X. Zeng, and L. Shang, ``The Power of Graph Signal Processing for Chip Placement Acceleration'', 
\emph{Proc. ICCAD}, 2024, pp. 1--8.

\bibitem{liu2025gsp-based}
Y. Liu, H. Zhou, J. Wang, F. Yang, X. Zeng, and L. Shang, ``Graph Signal Processing-Based Initialization for Chip Placement Acceleration'', 
\emph{IEEE Trans.CAD}, 2025.

\bibitem{eplace}
J. Lu, P. Chen, C.-C. Chang, L. Sha, D. J.-H. Huang, C.-C. Teng, and C.-K. Cheng, ``ePlace: Electrostatics-based placement using fast fourier transform and Nesterov's method'', 
\emph{TODAES}, 20(2), 2015, pp. 1--34.

\bibitem{nam2005ispd2005}
G.-J. Nam, C. J. Alpert, P. Villarrubia, B. Winter, and M. Yildiz, ``The ISPD2005 placement contest and benchmark suite'', 
\emph{Proc. ISPD}, 2005, pp. 216--220.

\bibitem{naylor2001non}
W. C. Naylor, R. Donelly, and L. Sha, ``Non-linear optimization system and method for wire length and delay optimization for an automatic electric circuit placer'', 
US Patent 6,301,693, 2001.

\bibitem{shuman2013emerging}
D. I. Shuman, S. K. Narang, P. Frossard, A. Ortega, and P. Vandergheynst, ``The emerging field of signal processing on graphs: Extending high-dimensional data analysis to networks and other irregular domains'', 
\emph{ISPM}, 30(3), 2013, pp. 83--98.

\bibitem{tunneling}
Natarajan Viswanathan, Gi-Joon Nam, Jarrod A. Roy, Zhuo Li, Charles J. Alpert, Shyam Ramji and Chris Chu, ``ITOP: Integrating timing optimization within placement'', 
\emph{Proc. ISPD}, 2010, pp. 83--90.

\bibitem{wu2019simplifying}
F. Wu, A. Souza, T. Zhang, C. Fifty, T. Yu, and K. Weinberger, ``Simplifying graph convolutional networks'', 
\emph{Proc. ICML}, 2019, pp. 6861--6871.

\bibitem{yao2025LLMevolution}
X. Yao, J. Jiang, Y. Zhao, P. Liao, Y. Lin, and B. Yu, ``Evolution of Optimization Algorithms for Global Placement via Large Language Models'', 
\emph{arXiv preprint arXiv:2504.17801}, 2025.

\bibitem{yutsis2014ispd}
V. Yutsis, I. S. Bustany, D. Chinnery, J. R. Shinnerl, and W.-H. Liu, ``ISPD 2014 benchmarks with sub-45nm technology rules for detailed-routing-driven placement'', 
\emph{Proc. ISPD}, 2014, pp. 161--168.

\bibitem{Agnesina2023AutoDMPAD}
A. Agnesina, P. Rajvanshi, T. Yang, G. Pradipta, A. Jiao, B. Keller, B. Khailany, and H. Ren, ``AutoDMP: Automated DREAMPlace-based Macro Placement'', 
\emph{Proc. ISPD}, 2023.

\bibitem{Hier-RTLMP}
A. B. {Kahng}, R. {Varadarajan}, and Z. {Wang}, ``Hier-RTLMP: A Hierarchical Automatic Macro Placer for Large-Scale Complex IP Blocks'', 
\emph{IEEE Trans.CAD}, 43(5), 2024, pp. 1552–1565.

\bibitem{10.1145/3377930.3389817}
Y. Ozaki, Y. Tanigaki, S. Watanabe, and M. Onishi, ``Multiobjective tree-structured parzen estimator for computationally expensive optimization problems'', 
\emph{Proc.GECC}, 2020, pp. 533–541.

\bibitem{weisstein2003gershgorin}
E. W. Weisstein, ``Gershgorin circle theorem'', 
\emph{https://mathworld. wolfram. com/}, 2003.

\bibitem{dramplace-github}
``The DREAMPlace Project'', 
\url{https://github.com/limbo018/DREAMPlace}.

\bibitem{ariane}
``Ariane {RISC-V CPU} repo'', 
\url{https://github.com/openhwgroup/cva6}.

\bibitem{bp_quad}
``{BlackParrot} repo'', 
\url{https://github.com/black-parrot/black-parrot}.

\bibitem{mempool_group}
``{MemPool} repo'', 
\url{https://github.com/pulp-platform/mempool}.

\end{thebibliography}


\begin{minipage}[ht]{0.48\textwidth}
    \begin{IEEEbiography}[{\includegraphics[width=1in,height=1.25in,clip,keepaspectratio]{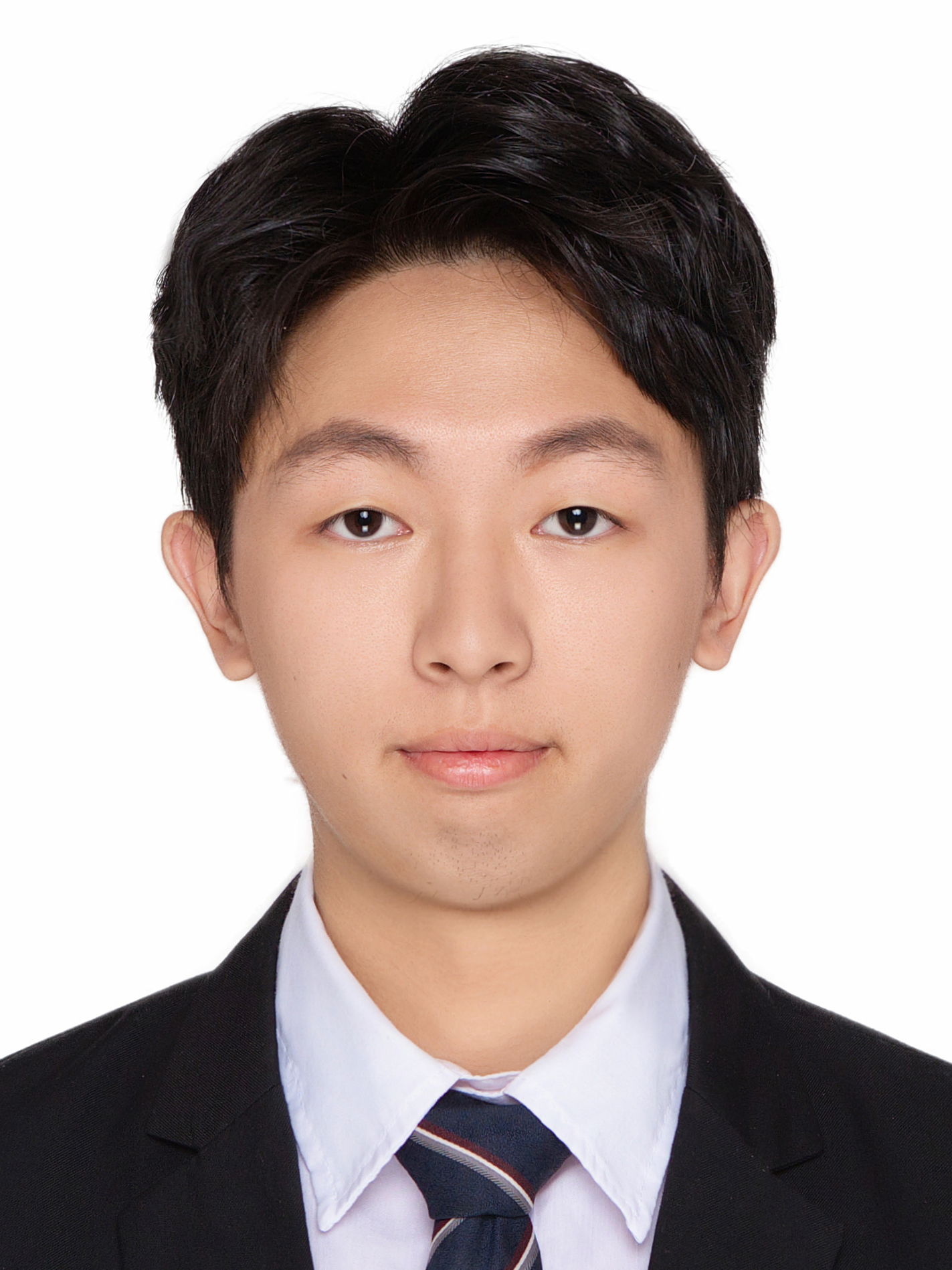}}]{Yuhao Ren}
        received the B.S. degree from Fudan University, Shanghai, China, in 2024. He is currently pursuing the Ph.D. degree in the College of Integrated Circuits and Micro-Nano Electronics at Fudan University. His current research interests include 3D IC physical design automation and spectral analysis in physical design.
    \end{IEEEbiography}
\end{minipage}
\hfill

\begin{minipage}[ht]{0.48\textwidth}
    \begin{IEEEbiography}[{\includegraphics[width=1in,height=1.25in,clip,keepaspectratio]{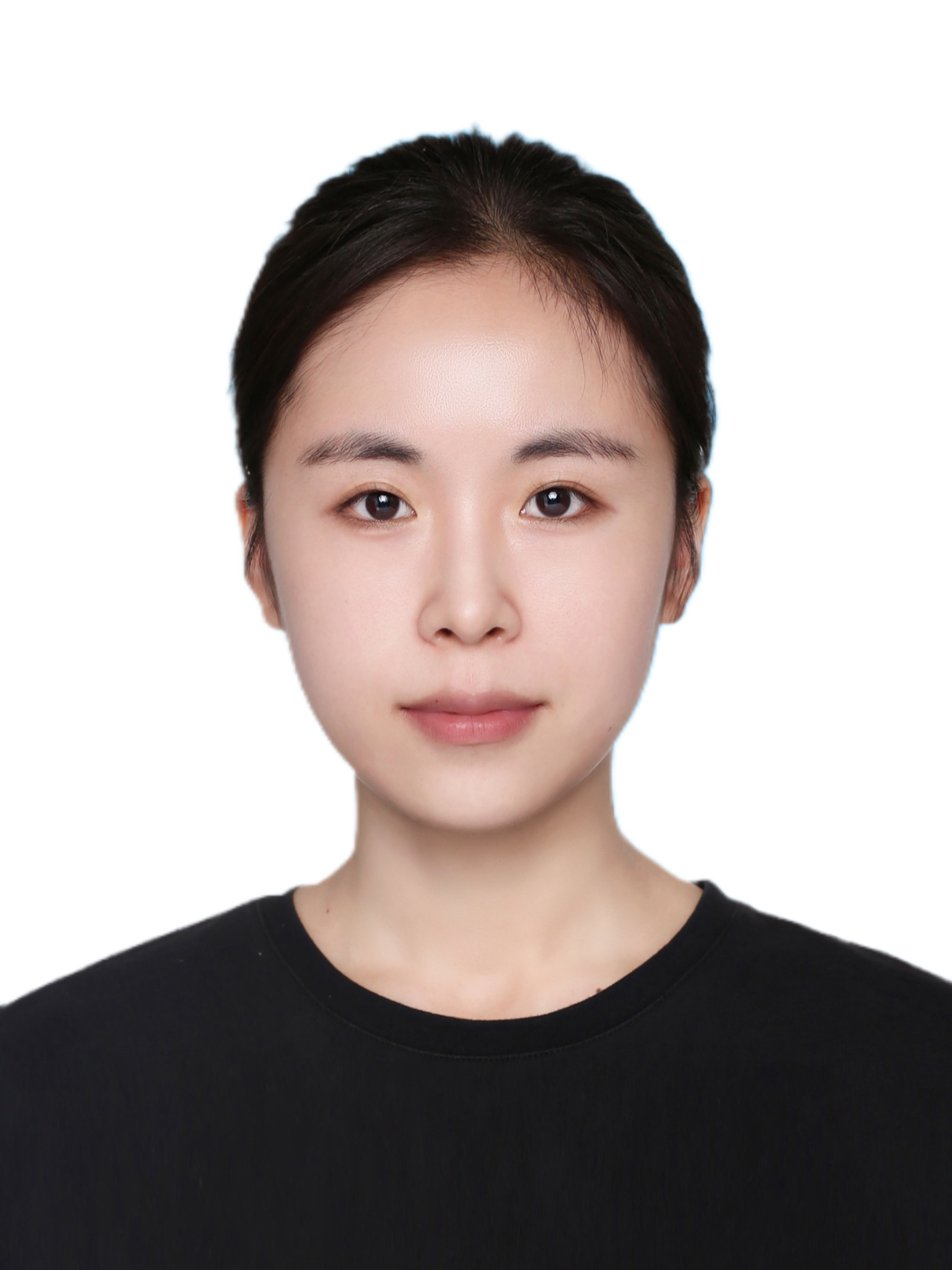}}]{Yiting Liu}         
    received the Ph.D. degree in computer science from Fudan University, Shanghai, China, in 2024. 
    She is currently a postdoctoral researcher at the University of California, San Diego. 
    Her current research interests include VLSI physical design, electronic design automation and machine learning.
    \end{IEEEbiography}
\end{minipage}
\hfill

\begin{minipage}[ht]{0.48\textwidth}
    \begin{IEEEbiography}[{\includegraphics[width=1in,height=1.25in,clip,keepaspectratio]{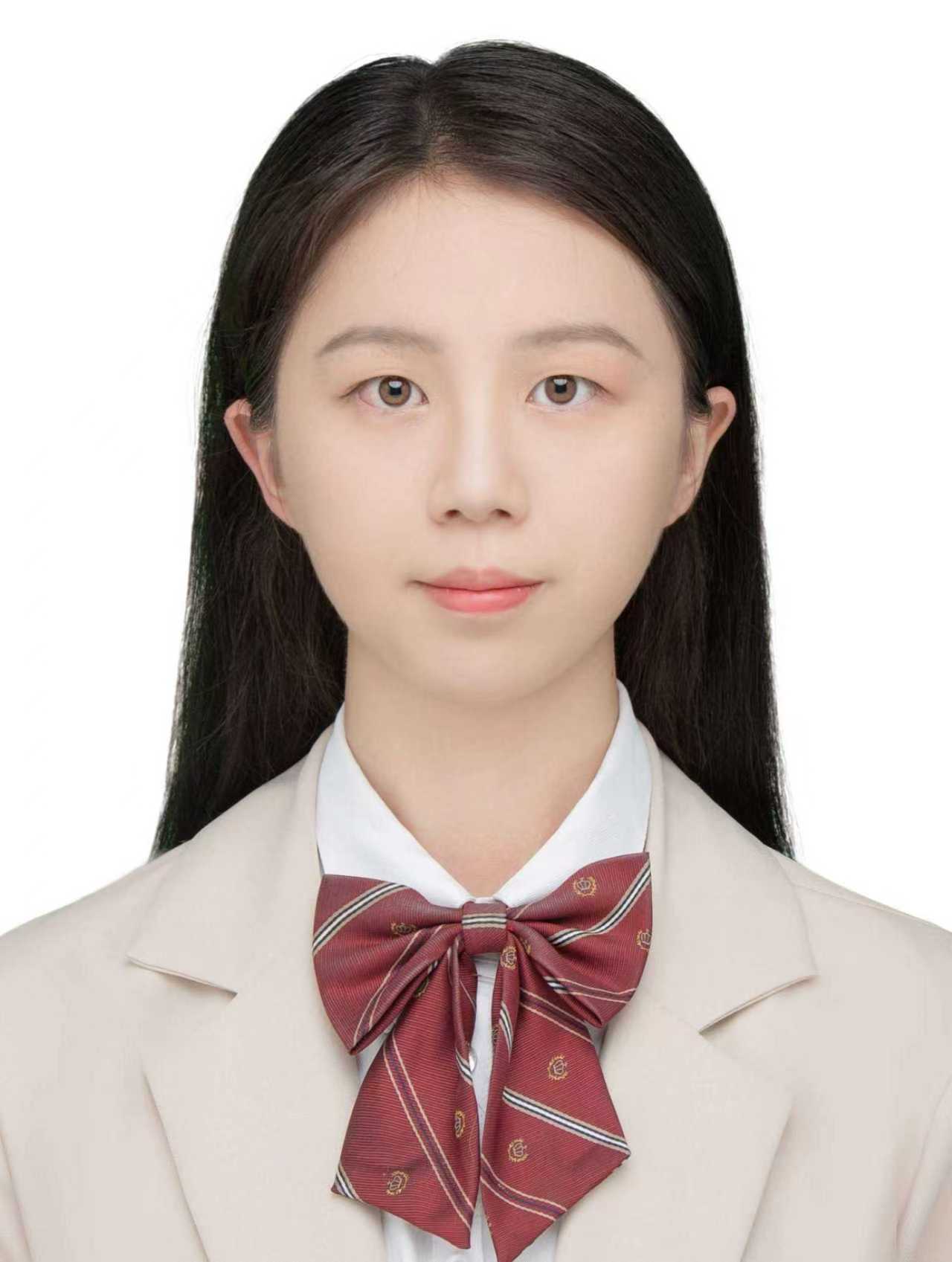}}]{Yanfei Zhou}
        received the B.E. degree in microelectronics science and technology from Wuhan University in 2024. She is currently pursuing a Ph.D. degree in electronic information in the College of Integrated Circuits and Micro-Nano Electronics at Fudan University. Her current research interests include VLSI physical design, focusing on algorithm, optimization, and machine learning techniques for advanced EDA.
    \end{IEEEbiography}
\end{minipage}
\hfill

\begin{minipage}[ht]{0.48\textwidth}
    \begin{IEEEbiography}[{\includegraphics[width=1in,height=1.25in,clip,keepaspectratio]{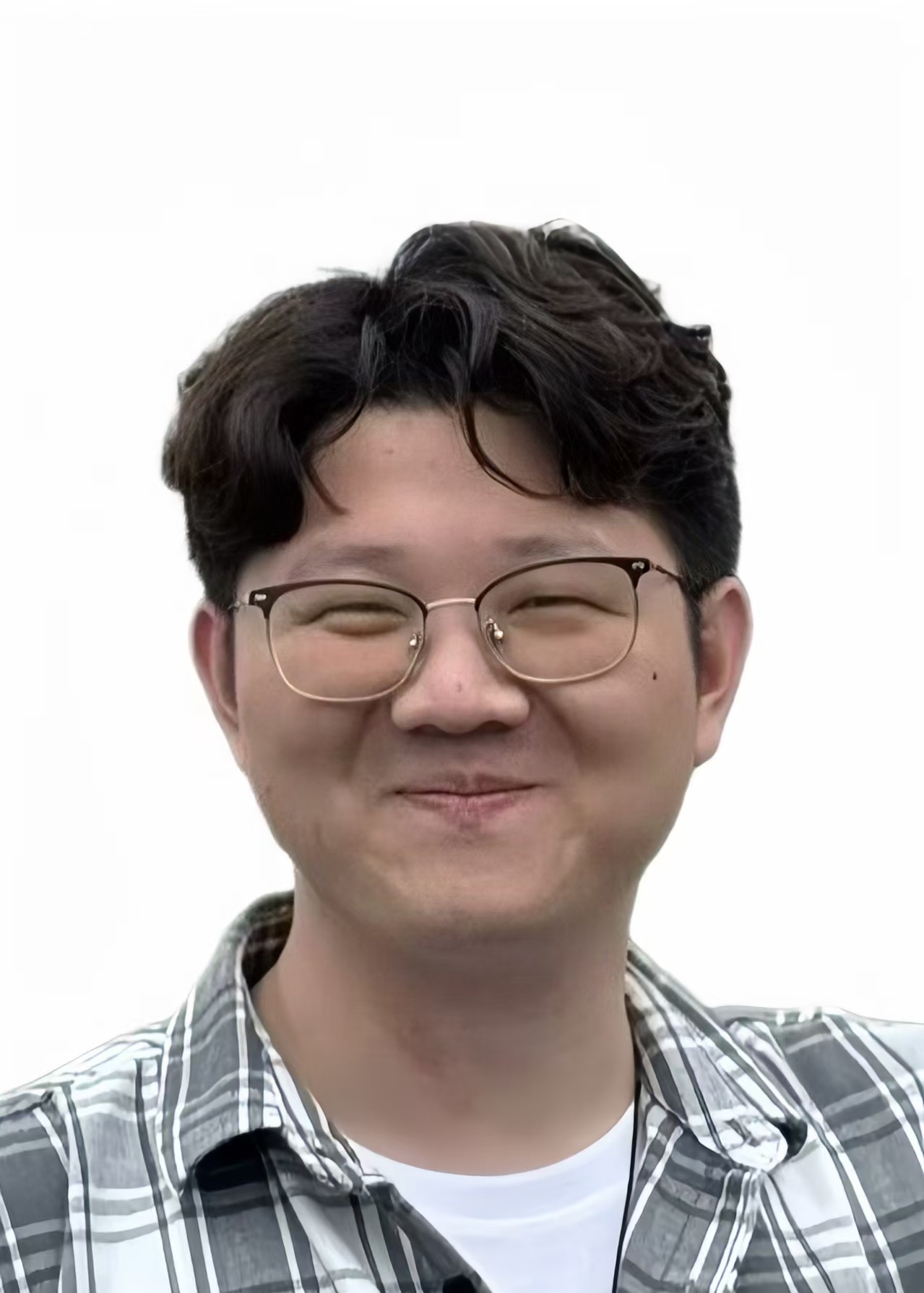}}]{Zhiyu Zheng}
        received the B.S. degree in Microelectronics Science and Engineering from Fudan University in 2024. He is currently pursuing the Doctor of Engineering (Eng.D.) degree in the College of Integrated Circuits and Micro-Nano Electronics at Fudan University. He actively contributes to the open-source community (https://github.com/shifengzhicheng). His research interests include ML4EDA, VLSI physical design algorithms, open-source EDA tool development, and 3D IC design methodologies.
    \end{IEEEbiography}
\end{minipage}
\hfill

\begin{minipage}[ht]{0.48\textwidth}
    \begin{IEEEbiography}[{\includegraphics[width=1in,height=1.25in,clip,keepaspectratio]{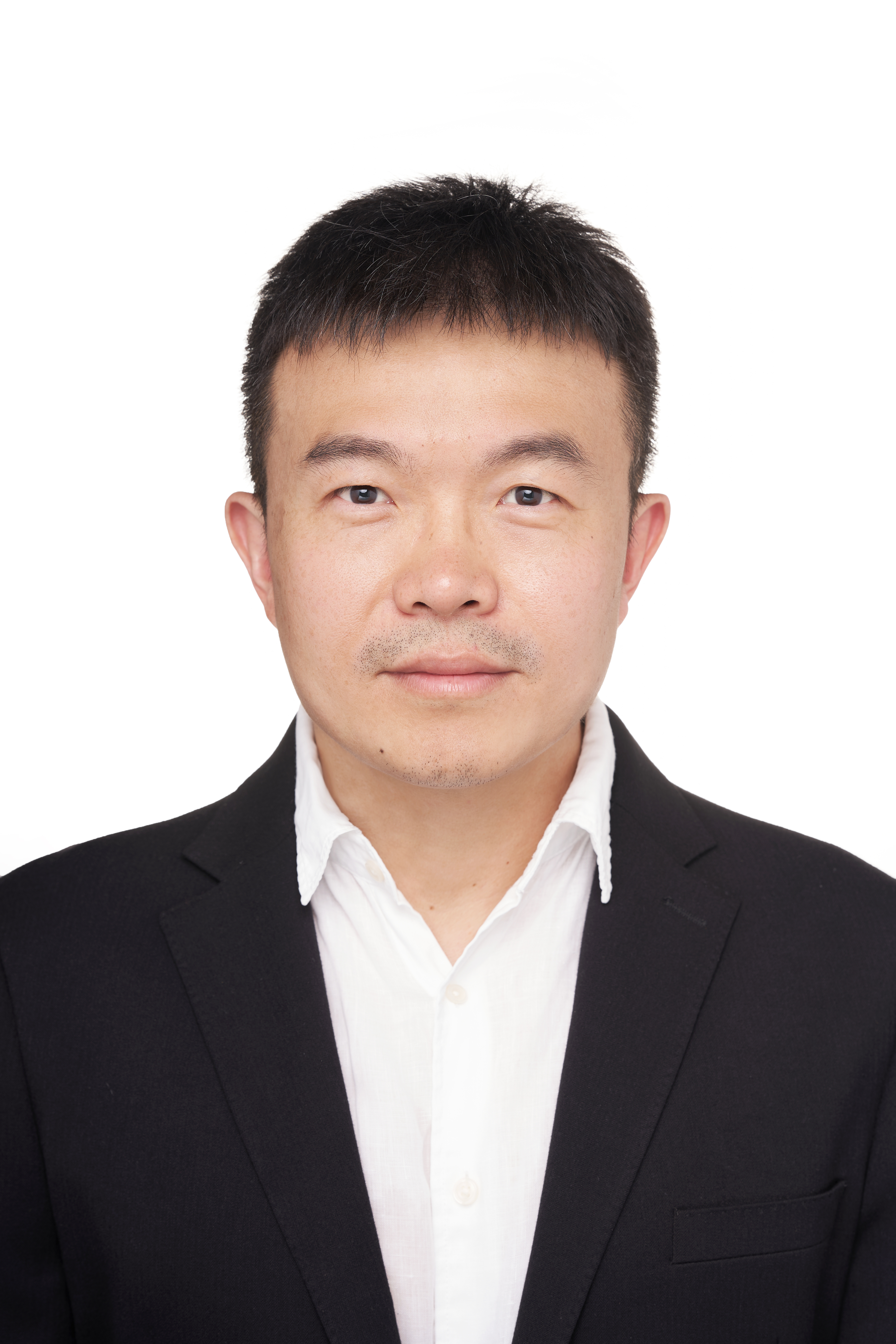}}]{Li Shang} is a Professor at the School of Computer Science at Fudan University. He received his Ph.D. degree from Princeton University. He was the Deputy Director and Chief Architect of Intel Labs China and an Associate Professor at the University of Colorado Boulder. His current research focuses on human-centered AGI algorithms, system software, and hardware. He has over 160 publications in human-centered computing, machine learning, computer systems, and electronic design automation, with multiple best paper awards and nominations and over 8000 citations. He was a recipient of the NSF Career Award.
    \end{IEEEbiography}
\end{minipage}
\hfill

\begin{minipage}[ht]{0.48\textwidth}
    \begin{IEEEbiography}[{\includegraphics[height=1.25in,clip,keepaspectratio]{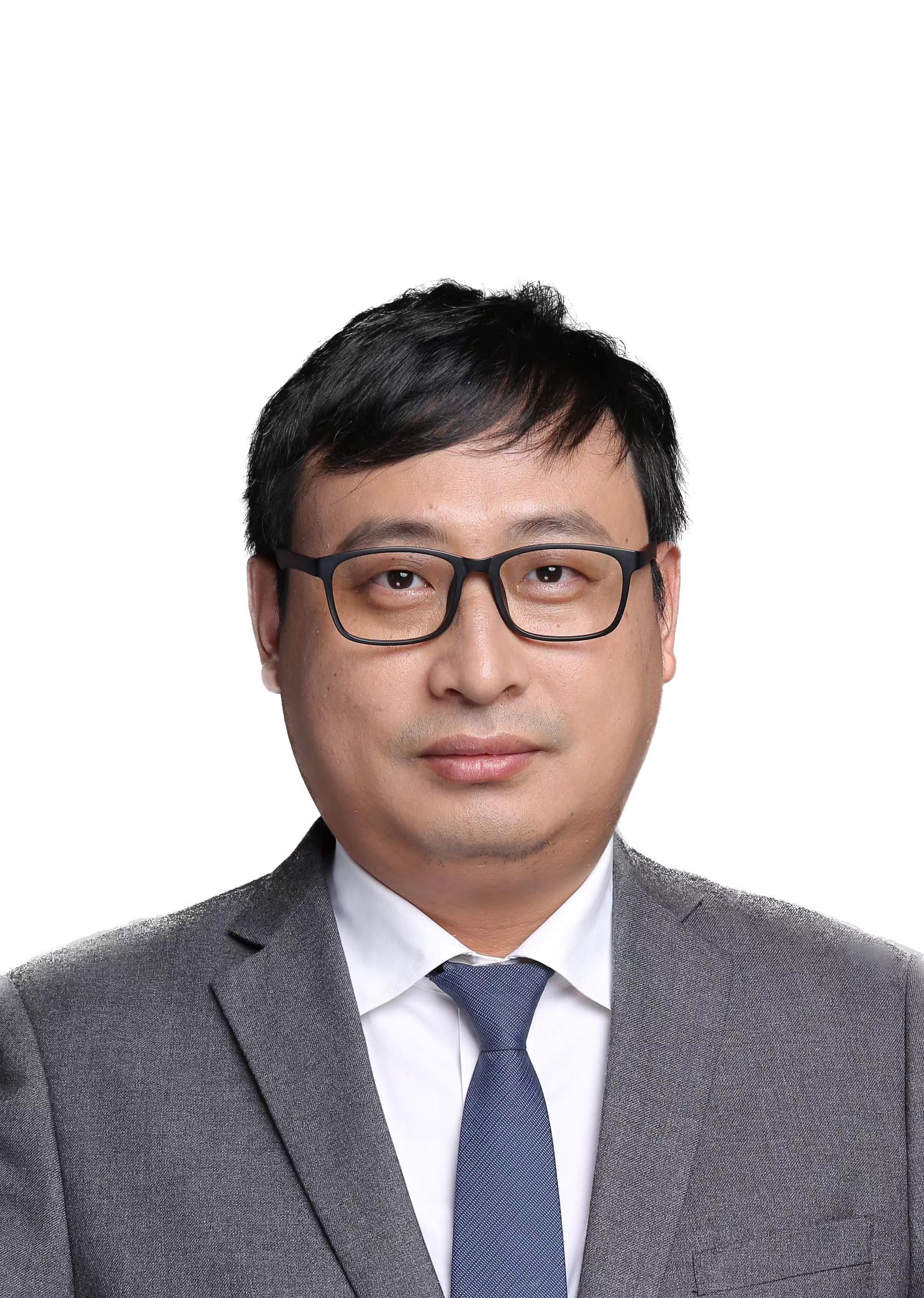}}]{Fan Yang} (Member, {IEEE}) 
        received the B.S. degree from Xi'an Jiaotong University, Xi'an, China, in 2003, and the Ph.D. degree from Fudan University, Shanghai, China, in 2008. He is currently a Full Professor with the School of Microelectronics, Fudan University. His research interests include model order reduction, circuit simulation, high-level synthesis, acceleration of artificial neural networks, and yield analysis and design for manufacturability.
    \end{IEEEbiography}
\end{minipage}
\hfill

\begin{minipage}[ht]{0.48\textwidth}
    \begin{IEEEbiography}[{\includegraphics[width=1in,height=1.25in,clip,keepaspectratio]{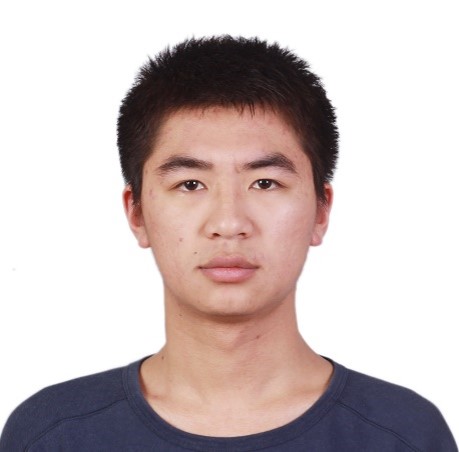}}]{Zhiang Wang} is currently an Assistant Professor with the College of Integrated Circuits and Nano-Micro Electronics, Fudan University.
    He received his Ph.D. degree in electrical and computer engineering from the University of California, San Diego in 2024. His current research interests include physical design, system technology co-optimization and machine learning for EDA.
    \end{IEEEbiography}
\end{minipage}
\hfill


\end{document}